\documentclass[showpacs,preprintnumbers,amssymb,twocolum,aps,reprint,superscriptaddress,showkeys,amsmath,floatfix]{revtex4-2}

\usepackage{graphicx}
\usepackage{dcolumn}
\usepackage[dvipsnames]{xcolor}
\usepackage{bm}
\usepackage{amsmath}
\usepackage{amssymb}
\usepackage{epsfig}
\usepackage{amsfonts}
\usepackage{lineno,hyperref}
\usepackage{array}
\usepackage{float}
\usepackage{microtype}
\usepackage{multirow}
\usepackage{adjustbox}
\usepackage[english]{babel}
\usepackage{epstopdf}
\usepackage{blindtext}
\usepackage{booktabs}
\usepackage{subcaption}
\usepackage[a4paper, total={6.7in, 10in}]{geometry}
\usepackage{appendix}

\def \be{\begin{equation}}
\def \ee{\end{equation}}
\def \ben{\begin{eqnarray}}
\def \een{\end{eqnarray}}

\def \n{\nonumber\\}

\def \La{\mathcal{L}}

\begin{document}

\title{Observational Constraints on \texorpdfstring{$f(Q,T)$}{} Gravity in the Presence of DBI-Essence Scalar Field}

\author{Mayur Mune} 
 \email{mayurmune1999@gmail.com}
\author{Praveen Kumar Dhankar} 
 \email{pkumar6743@gmail.com, praveen.dhankar@sitnagpur.siu.edu.in}
\affiliation{%
 \scriptsize Symbiosis Institute of Technology, Nagpur Campus, Symbiosis International (Deemed University), Pune 440008, Maharashtra, India.
}%

\author{Goutam Manna$^a$}
\email{goutammanna.pkc@gmail.com \\$^a$Corresponding author}
\affiliation{Department of Physics, Prabhat Kumar College, Contai, Purba Medinipur 721404, India} 
\affiliation{Institute of Astronomy, Space and Earth Science, Kolkata 700054, India}

\author{Safiqul Islam }
 \email{sislam@kfu.edu.sa}
\affiliation{
 \scriptsize Department of Mathematics and Statistics, College of Science, King Faisal University, P.O. Box 400, Al Ahsa 31982, Saudi Arabia.
}%

\author{Bidisha Samanta} 
 \email{samantabidisha@gmail.com }
\affiliation{
 \scriptsize Department of Mathematics, Sister Nivedita University, Kolkata, India.
}%

\author{Behnam Pourhassan} 
 \email{b.pourhassan@du.ac.ir}
\affiliation{
 \scriptsize School of Physics, Damghan University, Damghan, 3671641167,  Iran. \& Center for Theoretical Physics, Khazar University, 41 Mehseti Street, Baku, AZ1096, Azerbaijan.
}%

\date{\today}

\begin{abstract}
We investigate late-time cosmology in extended symmetric teleparallel gravity coupled to a Dirac-Born-Infeld (DBI) scalar field within $f(Q,T)$ gravity, where $Q$ is the non-metricity scalar and $T$ is the trace of the matter energy-momentum tensor. Working on a spatially flat Friedmann--Lema\^itre--Robertson--Walker background and treating the cosmic medium as an effective perfect fluid, we derive the background field equations for $f(Q,T)+\mathrm{DBI}$ gravity and obtain analytic solutions for the linear choice $f(Q,T)=\alpha Q+\beta T$. We then constrain the model parameters with a Markov Chain Monte Carlo analysis using Hubble-rate data, DESI BAO (DR2) measurements, and the Pantheon+SHOES Type~Ia supernova sample. The joint posteriors (Tables~\ref{tab:placeholder11} and \ref{tab:placeholder}) are broadly consistent with current late-time constraints and allow a direct comparison with $\Lambda$CDM, quantifying the departures driven by the $\beta T$ coupling and the DBI sector. Although the model does not reproduce every observational feature exactly, it provides a statistically viable alternative avenue to the standard paradigm and a useful framework for exploring potential remedies to existing tensions, including the $H_0$ discrepancy, without claiming a definitive resolution.
\end{abstract}

\maketitle


\section{Introduction}
Baryon Acoustic Oscillations (BAO) \cite{5,6}, Cosmic Microwave Background (CMB) anisotropies \cite{7,8}, large-scale structure (LSS) surveys \cite{3,4}, and Type~Ia supernova (SNe~Ia) observations \cite{1,2} collectively support the late-time accelerated expansion of the universe. According to recent Planck constraints, the present cosmic energy budget is dominated by dark energy (DE), contributing $\sim 70\%$, with dark matter and baryons accounting for $\sim 25\%$ and $\sim 5\%$, respectively \cite{9}. While the $\Lambda$CDM paradigm provides an excellent fit to a wide range of observations, it is challenged by conceptual issues such as the cosmological constant and coincidence problems \cite{Weinberg1989, Carroll2001}, which has motivated many extensions. On the matter side, dynamical DE models including quintessence \cite{10,11}, k-essence \cite{12,13}, DBI-essence \cite{14}, and Chaplygin gas scenarios \cite{15,16,43,Praveen2025} have been proposed; in particular, non-canonical kinetic terms (k-essence/DBI) can lead to richer late-time dynamics and attractor behavior \cite{17,18}. In parallel, observational data combined with Bayesian/MCMC techniques have been widely used to test departures from GR in frameworks such as $f(G)$, $f(T)$, Gauss--Bonnet, and modified $f(R)$ gravity \cite{P1,T1,D1,D2,D3,D4,D5}.

From the geometric perspective, non-Riemannian extensions of GR, notably symmetric teleparallel gravity, encode gravitation in the non-metricity scalar $Q$ \cite{Nester1998, BeltranJimenez2018}. Generalizing $Q$ to $f(Q)$ can account for late-time acceleration through modified gravitational dynamics without an explicit dark-energy fluid \cite{Jimenez2018, Lazkoz2019}. An additional extension, $f(Q,T)$ gravity, allows the Lagrangian to depend on both $Q$ and the trace of the matter energy-momentum tensor $T$, introducing a direct matter-geometry coupling \cite{Xu2019}. This coupling generically leads to $\nabla_\mu T^{\mu\nu}\neq 0$, which can be viewed as an effective energy-momentum exchange between matter and geometry; in cosmology it can be interpreted in terms of particle production and formulated consistently within the thermodynamics of open systems \cite{Harko2011, Prigogine1989}. These features make $f(Q,T)$ gravity a flexible setting for exploring the joint impact of modified geometry and effective matter creation on cosmic evolution.

On the other hand, scalar fields have long served as a minimal and versatile ingredient for modeling both primordial inflation and the current accelerated expansion of the universe \cite{Guth1981, Linde1982, AlbrechtSteinhardt1982, CopelandSamiTsujikawa2006}. Canonical scenarios, including quintessence, offer a simple dynamical dark-energy candidate; however, their standard kinetic structure can be too restrictive to capture the full phenomenology suggested by observations, which motivates generalized (non-canonical) scalar-field models \cite{RatraPeebles1988, Caldwell1998}. A widely studied class is k-essence, where the Lagrangian depends nonlinearly on the kinetic term $X\equiv-\tfrac12\nabla_\mu\phi\nabla^\mu\phi$ \cite{ArmendarizPicon2000, Garriga1999, Chiba2000, ArmendarizPicon2001}. Dirac-Born-Infeld (DBI) or DBI-essence models can be viewed as a particular (string-motivated) realization within this broader k-essence framework, characterized by a square-root kinetic structure \cite{Silverstein2004, Alishahiha2004}. These kinetic self-interactions typically lead to a nontrivial effective sound speed and can substantially affect both the background expansion and the evolution of cosmological perturbations; in the DBI case, the associated speed limit for field motion can support inflationary dynamics and may produce distinctive signatures such as enhanced primordial non-Gaussianities \cite{Maldacena2003, Chen2005}. More generally, k-essence-type theories can drive accelerated expansion without requiring an exceptionally flat potential and provide viable realizations of both inflation and dynamical dark energy \cite{Chiba2000, ArmendarizPicon2001, Scherrer2004}. It is also worth noting that k-essence/DBI-essence models can be embedded within generalized scalar-tensor theories, including Horndeski gravity \cite{Horndeski, Deffayet, Kobayashi}. \\

It is therefore natural to consider a unified setup in which modified gravity is coupled to a non-canonical scalar field. The motivation is twofold. Geometrically, $f(Q,T)$ theories modify the gravitational sector that drives cosmic expansion and, through their explicit dependence on $T$, introduce matter-geometry couplings that typically imply $\nabla_\mu T^{\mu\nu}\neq 0$. This non-conservation can be interpreted as an effective exchange of energy-momentum between matter and gravity. From the field-theory perspective, k-essence/DBI-type scalars add genuine dynamical freedom, most notably through a variable effective sound speed, which can support both inflationary and late-time accelerated phases and may reduce the level of fine-tuning compared with purely canonical models. Together, these ingredients provide a well-defined arena to investigate how modified geometry and nonlinear kinetic effects shape the background expansion and the growth of perturbations.

From a theoretical standpoint, this combination is particularly attractive. Since the non-metricity scalar $Q$ is constructed from first derivatives of the metric, many $f(Q,T)$ models lead to second-order field equations, thereby avoiding the higher-derivative instabilities that can arise in generic curvature-based extensions. Similarly, DBI and more general k-essence actions depend on $\phi$ and its kinetic term $X$ and yield second-order equations of motion, capturing nonlinear kinetic effects without introducing additional ghost-like degrees of freedom. Phenomenologically, the $T$-dependent coupling can mediate energy transfer among the scalar sector, standard matter, and the gravitational field, often interpretable as effective particle production, and can modify the expansion rate, the effective equation of state, and the evolution of cosmological perturbations. As a result, the framework can lead to potentially testable departures from the standard picture of structure formation.\\

The main objective of this work is to study the cosmological implications of an $f(Q,T)$ gravity model supplemented by a non-canonical DBI scalar field. Starting from the total action, we derive the modified field equations and then specialize to a spatially homogeneous and isotropic FLRW background to obtain the corresponding modified Friedmann equations. We use these equations to quantify how the DBI kinetic structure and the explicit $T$-dependence modify the background expansion history.

The same framework also offers a unified arena to discuss both an early inflationary stage and late-time acceleration. Because the scalar sector is non-canonical, the perturbation dynamics can be richer than in canonical models, with modified sound speed and stability conditions that may leave observational signatures in the Hubble expansion, the growth of large-scale structure, and the propagation of gravitational waves \cite{Bhunia}.

Overall, the combination of $f(Q,T)$ gravity with DBI/k-essence-type scalars provides a physically motivated and extensible setting to explore the interplay between modified geometric dynamics, nonlinear kinetic physics, and effective matter-creation effects, with applications ranging from inflationary cosmology to the late Universe.\\

In this work, we construct an $f(Q,T)+\mathrm{DBI}$ cosmological scenario on a spatially flat FLRW background and derive the associated modified Friedmann equations. For the representative choice $f(Q,T)=\alpha Q+\beta T$, we obtain analytic solutions for the background evolution and confront the model with recent expansion-history observations through a Markov chain Monte Carlo (MCMC) analysis of $H(z)$ data, DESI BAO measurements, and the Pantheon+SHOES Type~Ia supernova compilation. The joint constraints are then used to test the model's observational viability and to compare its predictions with those of the standard $\Lambda$CDM cosmology, with particular attention to its potential relevance for late-time tensions such as the $H_0$ discrepancy. We do not address additional aspects, for example, a full perturbation-level analysis, detailed large-scale-structure constraints, or early-universe inflationary model building, which are beyond the scope of the present study.\\

Throughout this work, we use the metric signature $(-,+,+,+)$. We set $c=1$ in the theoretical equations and restore $c$ in all observational (distance-measure) formulas; since parameter estimation uses only the observational sector, this causes no ambiguity. We also set the reduced Planck mass $M_{\rm Pl}=1$.  \\

The paper is organized as follows. In Sec.~\eqref{S2}, we introduce the $f(Q,T)+\mathrm{DBI}$ framework, derive the background field equations, and present the key evolution equations governing the cosmological dynamics. Section~\eqref{S3} describes the observational and statistical methodology used to constrain the model parameters: we outline the Bayesian MCMC approach and summarize the datasets employed in our analysis, namely $H(z)$ measurements, DESI BAO data, and the Pantheon+SHOES Type~Ia supernova compilation. In Sec.~\eqref{S4}, we report the resulting parameter constraints (including best-fit values and confidence contours) and compare the proposed scenario with the standard $\Lambda\mathrm{CDM}$ and $\omega\mathrm{CDM}$ models using model-selection statistics such as $\chi^2$, $\chi^2_{\rm red}$, AIC, BIC, and DIC. Finally, Sec.~\eqref{S5} summarizes our main conclusions.

\section{Dynamics of $f(Q,T)+\mathrm{DBI}$ gravity}
\label{S2}

We start from the action
\ben
S = \int d^4x\sqrt{-g}\Big(\frac{1}{2}f(Q,T)+\La_{\rm DBI}(\phi,X)+\La_{\rm m}\Big),
\label{1}
\een
where $f(Q,T)$ is a generic function of the non-metricity scalar $Q$ and the trace of the energy-momentum tensor $T$. The term $\La_{\rm DBI}$ is a non-canonical Dirac-Born-Infeld (DBI) Lagrangian for a scalar field $\phi$ with kinetic term $X$, $\La_{\rm m}$ denotes the matter Lagrangian, and $g\equiv\det(g_{\mu\nu})$. We work in units $8\pi G=1$.

The action \eqref{1} combines three ingredients: modified gravity based on non-metricity, a non-canonical scalar sector, and ordinary matter. In the symmetric teleparallel description, gravity is encoded in $Q$ rather than spacetime curvature, and suitable choices of $f(Q,T)$ can reproduce late-time acceleration without introducing an explicit cosmological constant \cite{Nester1998, BeltranJimenez2018, Jimenez2018, Lazkoz2019}. The dependence on $T$ produces an explicit coupling between matter and geometry; as a result, the matter energy-momentum tensor is not conserved in general ($\nabla_{\mu}T^{\mu\nu}\neq 0$). At the level of cosmology, this can be viewed as an effective exchange of energy between the gravitational and matter sectors (often interpreted as particle production in an open system) \cite{Xu2019, Harko2011, Prigogine1989}.

The DBI term is motivated by string theory and features a characteristic square-root structure that bounds the field's effective ``velocity'' \cite{Silverstein2004, Alishahiha2004}. Such non-canonical dynamics can generate a wide range of cosmological histories, from inflationary phases to late-time acceleration, and may be regarded as a particular realization of DBI-essence (k-essence) where the kinetic structure controls the evolution \cite{ArmendarizPicon2000, Garriga1999, Chiba2000}. The inclusion of $\La_{\rm m}$ ensures that the model contains the standard matter sources.

Note that the DBI field is minimally coupled in the Lagrangian (it does not appear explicitly in $f$), but the $T$-dependence of $f(Q,T)$ still introduces a non-minimal interaction between geometry and the matter sector through the field equations. For instance, in the simple choice $f(Q,T)=\alpha Q+\beta T$, the $\beta T$ term depends on $\La_{\rm m}$ and is therefore responsible for the matter-geometry coupling, while the DBI sector remains separate in the action but contributes to the total energy-momentum tensor that drives the background dynamics.\\

In symmetric teleparallel gravity, the gravitational interaction is attributed to spacetime non-metricity rather than curvature or torsion. The non-metricity tensor is defined by \cite{Xu2019, Heisenberg}
\ben
Q_{\alpha\mu\nu}=\nabla_{\alpha}g_{\mu\nu},
\label{2}
\een
which describes how the metric tensor changes under parallel transport. Geometrically, non-metricity allows the norm of a vector to vary along a curve. In the symmetric teleparallel framework one works with a connection for which curvature and torsion vanish, so that the gravitational dynamics are fully encoded in the non-metricity scalar $Q$.

\noindent
The non-metricity scalar $Q$ is defined as 
\ben
Q\equiv g^{\mu\nu}\Big(-L^\alpha_{\beta\mu}L^\beta_{\nu\alpha}+L^\alpha_{\beta\alpha}L^\beta_{\mu\nu}\Big),
\label{3}
\een
where $L^\alpha_{\beta\gamma}$ denotes the disformation tensor defined as
\ben
L^\alpha_{\beta\gamma}=-\frac{1}{2}g^{\alpha\lambda}(\nabla_{\gamma}g_{\beta \lambda}+\nabla_\beta g_{\lambda \gamma}-\nabla_\lambda g_{\beta \gamma}).
\label{4}
\een
\noindent
The traces of the non-metricity tensor are defined as follows: 
\ben
Q_\alpha \equiv Q_{\alpha}{}^{\mu}{}_{\mu}, 
\qquad 
\tilde{Q}_\alpha \equiv Q^{\mu}{}_{\alpha\mu}.
\label{5}
\een
Utilizing these quantities, one can formulate the superpotential tensor (or non-metricity conjugate) as \cite{Xu2019}
\ben
&&P^{\alpha}{}_{\mu \nu}\equiv\frac{1}{4}\Bigg[-Q^{\alpha}{}_{\mu \nu}+2Q_{(\mu}{}^\alpha{}_{\nu)}+Q^\alpha g_{\mu \nu}\n &&
-\tilde{Q}^{\alpha}g_{\mu \nu}-\delta^\alpha{}_{(\mu}Q{}_{\nu)}\Bigg]\n &&=-\frac{1}{2}L^\alpha{}_{\mu \nu}+\frac{1}{4}\left(Q^\alpha-\tilde{Q}^\alpha\right) g_{\mu \nu}-\frac{1}{4}\delta^\alpha{}_{(\mu}Q{}_{\nu)}
\label{6}
\een
According to the above definition, the non-metricity scalar can be concisely represented as 
\ben 
Q&&=-Q_{\alpha \mu \nu}P^{\alpha \mu \nu } \n &&= -\frac{1}{4}(-Q^{\alpha \nu \rho}Q_{\alpha \nu \rho}+2Q^{\alpha \nu \rho}Q_{\alpha \nu \rho}-2Q^\rho \tilde{Q}^\alpha +Q^\rho Q_\rho)\n
\label{7}
\een
The non-metricity scalar $Q$ plays a role analogous to the Ricci scalar $R$ in curvature-based formulations of gravity. In the symmetric teleparallel equivalent of general relativity (STEGR), also known as (coincident) general relativity, the action built from $Q$ is dynamically equivalent to GR and reproduces the Einstein field equations \cite{Nester1998, BeltranJimenez2018}. Modified scenarios follow by promoting $Q$ to a generic function, such as $f(Q)$ or by allowing an explicit dependence on the matter trace through $f(Q,T)$ \cite{Jimenez2018, Lazkoz2019, Xu2019}. Such models have been widely employed in cosmology to investigate late-time acceleration and possible deviations from general relativity \cite{Jimenez2018, Lazkoz2019}.\\

Applying the variation principle to the action \eqref{1} with respect to $g_{\mu\nu}$, we found the modified field equations as \cite{Pal, Mandal, gm3, gm4, Ganguly, Vikman, Martin, Xu2019, Heisenberg} (\emph{For full derivation, see Appendix \ref{A}})
\ben
&&\frac{-2}{\sqrt{-g}}\nabla^\alpha (f_Q\sqrt{-g}~P^{\alpha}_{\mu\nu})-\frac{1}{2}f g_{\mu \nu}+f_T(T_{\mu\nu}+\Theta_{\mu\nu})\n &&-f_Q(P_{\mu \alpha \beta }Q^{\alpha\beta}_\nu-2Q^{\alpha\beta}{}_\mu P_{\alpha\beta\nu})= T^{(m)}_{\mu\nu}+T^{\rm DBI}_{\mu\nu},\n
\label{8}
\een 
where 
\ben
T^{(\rm m)}_{\mu \nu}=-\frac{2}{\sqrt{-g}}\frac{\delta (\sqrt{-g}\mathcal{L}_{\rm m})}{\delta g_{\mu\nu}}~;~\Theta_{\mu\nu}=g^{\alpha\beta}\frac{\delta T_{\alpha\beta}}{\delta g_{\mu\nu}}\n
\label{9}
\een 
with $\delta T= \delta(T_{\mu\nu}g^{\mu\nu})=(T_{\mu\nu}+\Theta_{\mu\nu})\delta g^{\mu \nu}$ and 
\ben
T^{\rm DBI}_{\mu \nu}&&=-\frac{2}{\sqrt{-g}} \frac{\delta (\sqrt{-g}\La_{\rm DBI})}{\delta g^{\mu \nu}}\n && \equiv  g_{\mu\nu}\La_{\rm DBI}-(\La_{\rm DBI})_X \nabla_\mu \phi \nabla_\nu \phi,
\label{10}
\een
where $(\La_{\rm DBI})_X=\frac{\partial \La_{\rm DBI}}{\partial X}$.\\

In our study, we consider the non-canonical Lagrangian associated with the Dirac-Born-Infeld (DBI) scalar field given by \cite{Vikman, Martin, Mandal, Ganguly}:

\ben
\La_{\rm DBI}(\phi,X)
=
- f(\phi)\Big(\sqrt{1-\frac{2X}{f(\phi)}}-1\Big)-V(\phi),
\label{11}
\een
where $X$ is defined as
$X\equiv -\frac{1}{2}\,g^{\mu\nu}\partial_{\mu}\phi\,\partial_{\nu}\phi$, which corresponds to the canonical kinetic term of the DBI Lagrangian. Here, $f(\phi)$ denotes the warp (brane-tension) factor, referred to as the tension scalar, whereas $V(\phi)$ is the scalar potential. Throughout this work we assume a homogeneous scalar field, $\phi(r,t)\equiv \phi(t)$, as commonly adopted in DBI-essence (k-essence) cosmology \cite{gm3,gm4}. We further define the DBI Lorentz factor as \cite{Ganguly,Pal}
\ben
\gamma\equiv\frac{1}{\sqrt{1-\dfrac{2X}{f(\phi)}}}.
\label{12}
\een
The Lorentz factor depends on both the kinetic term ($2X$) and the tension scalar $f(\phi)$. Requiring the square root in Eq.~\eqref{12} to be real and positive imposes $\frac{2X}{f(\phi)}<1$, which implies $\gamma>1$. Hence, $f(\phi)$ controls the admissible range of the field velocity, playing a role analogous to the speed-of-light bound in special relativity.\\

Varying the DBI action in Eq.~\eqref{1}, with the DBI Lagrangian \eqref{11}, with respect to the scalar field $\phi$ and the metric $g_{\mu\nu}$ in a spatially flat FLRW background yields the scalar-field equation of motion (EoM) and the associated energy--momentum tensor $\big(T^{\mu}{}_{\nu(\phi)}\big)$, given respectively by

\ben
&&\Ddot{\phi}- \frac{3 f'(\phi)}{2 f(\phi)}\dot{\phi}^2+ f'(\phi)+\frac{3H}{\gamma^2}\dot{\phi}\n &&+ \frac{1}{\gamma^3}[V'(\phi)-f'(\phi)]=0 
\label{13}
\een
and
\ben
T_{\mu\nu(\phi)} = (\rho_{\phi} + P_{\phi}) u_{\mu}u_{\nu}+ P_{\phi}\,g_{\mu\nu} ,
\label{14}
\een
with 
\ben
&& u_{\mu}=\frac{\partial_{\mu} \phi}{\sqrt{2X}} \quad ;\quad  u_{\mu}u^{\mu}=-1. 
\label{15}
\een
In Eq.~\eqref{13}, a `prime' denotes differentiation with respect to the scalar field $\phi$, whereas an `overdot' denotes differentiation with respect to the cosmic time $t$. In Eq.~\eqref{14}, $\rho_{\phi}$ and $P_{\phi}$ are, respectively, the energy density and pressure of the scalar-field sector. The Hubble parameter is defined as $H\equiv \dot a/a$, where $a\equiv a(t)$ is the scale factor. Accordingly, the energy density and pressure of the DBI scalar field are given by \cite{Martin,Ganguly}:

\ben
&&\rho_{\phi} = (\gamma - 1) f(\phi) +V(\phi)  ;\n && P_{\phi} = \big(\frac{\gamma -1}{\gamma}\big)f(\phi) - V(\phi)
\label{16}
\een

To characterize the dynamical evolution of the DBI scalar field, we introduce the \emph{equation-of-state parameter} $w_{\phi}$ and the (adiabatic) sound speed $c_{s(\phi)}$, defined as :

\ben
w_{\phi} && = \frac{P_{\phi}}{\rho_{\phi}}=\frac{ \big(\frac{\gamma -1}{\gamma}\big)f(\phi) - V(\phi)}{(\gamma - 1) f(\phi) +V(\phi)} \n  && =  \frac{ \big(\frac{\gamma -1}{\gamma}\big)\frac{f(\phi)}{V(\phi)} -1}{(\gamma - 1) \frac{f(\phi)}{V(\phi)} + 1}\equiv  \frac{ \big(\frac{\gamma -1}{\gamma}\big)r(\phi) -1}{(\gamma - 1) r(\phi) + 1} \n  
c_{s(\phi)}^2 && = \frac{\frac{\partial P_{\phi}}{\partial X}}{\frac{\partial \rho_{\phi}}{\partial X}} = \frac{1}{\gamma^2}.
\label{17}
\een
The Lorentz-like factor $\gamma$ plays a central role in the DBI dynamics through its connection to $w_{\phi}$ and $c_{s(\phi)}^{2}$. Following \cite{Ganguly}, we introduce the dimensionless ratio
$r(\phi)\equiv \dfrac{f(\phi)}{V(\phi)}$,
which is well defined since $f(\phi)$ and $V(\phi)$ share the same mass dimension. From the structure of Eq.~\eqref{17}, one sees that $r(\phi)$ largely governs the overall behaviour of $w_{\phi}$. Moreover, because $\gamma>1$ in DBI models, the factor $(\gamma-1)/\gamma$ satisfies $0<(\gamma-1)/\gamma<1$. For the DBI field to exhibit dark-energy-like behaviour compatible with cosmic acceleration, the equation of state requires
$r(\phi)\,\frac{\gamma-1}{\gamma}<1$,
which ensures sufficiently negative pressure. This motivates restricting the functional forms of $f(\phi)$ and $V(\phi)$ such that the above condition holds throughout the cosmological evolution of interest.
\\

\subsection{Cosmological Equations for the Action \texorpdfstring{\eqref{1}}{eq1}}

We now derive the modified Friedmann equations governing cosmic evolution in $f(Q,T)$ gravity in the presence of a DBI scalar field. We adopt a spatially flat, homogeneous, and isotropic Friedmann--Lema\^itre--Robertson--Walker (FLRW) spacetime with line element
\ben
 ds^2= -dt^2+a^2(t)\sum_{i=1}^{3}(dx^{i})^{2},
\label{18}
\een
for which the non-metricity scalar reduces to
\ben
Q=6H^2.
\label{19}
\een
We further introduce the notations \cite{Xu2019} 
\ben
F\equiv f_{Q}=\frac{\partial f}{\partial Q},\qquad \tilde{G}\equiv f_{T}=\frac{\partial f}{\partial T}.
\label{20}
\een

In a spatially homogeneous and isotropic FLRW background, the microscopic properties of ordinary matter and dark matter can be consistently coarse-grained into an effective fluid description. The perfect-fluid approximation is therefore a natural and minimal choice: it captures the dominant, isotropic part of the stress-energy (with a single equation-of-state relation), it is adequate for pressureless cold dark matter and standard cosmological components, and it avoids introducing model-dependent dissipative effects (viscosity, heat flow, or anisotropic stresses) that are subleading at the background level. Accordingly, we model the matter sector as a perfect fluid with energy-momentum tensor
\ben
T_{\mu}^{\nu} = {\rm diag}\,(-\rho, P, P, P),
\label{21}
\een
where $\rho$ and $P$ denote the total energy density and pressure, respectively. Then the tensor $\Theta_{\mu}^{\nu}$ reads
$\Theta_{\mu}^{\nu}=-2T_{\mu}^{\nu}+P\delta_{\mu}^{\nu}={\rm diag}\,(2\rho+P,-P,-P,-P)$.
The total energy density and pressure in our setup receive contributions from both the matter sector and the DBI scalar field,
\ben
\rho=\rho_{\rm m}+\rho_{\phi},\qquad
P=P_{\rm m}+P_{\phi}.
\label{22}
\een

The modified Friedmann equations can be expressed using the field equations \eqref{8} as follows: The first and second Friedmann equations are expressed as 
\ben
\frac{f}{2}-6FH^2=\rho+f_T(\rho+P)
\label{23}
\een
and 
\ben
\frac{f}{2}-2[\dot{F}H+F\dot{H}+3H^2F]=-P
\label{24}
\een
where $f=f(Q,T)$. Combining the above two Eqs. \eqref{23} and \eqref{24}, we get the following evolution equation for the Hubble function as
\ben
\dot{H}+\frac{\dot{F}}{F}H=\frac{1}{2F}(1+f_{T})(\rho+P).
\label{25}
\een
To facilitate comparison with standard cosmology, the modified cosmic evolution equations can be expressed in a manner akin to the Friedmann equations of general relativity by introducing an effective energy density $\rho_{\rm eff}$ and an effective pressure $P_{\rm eff}$, respectively, as
\ben
3H^2=\rho_{\rm eff}=\frac{f}{4F}-\frac{1}{2F}[(1+f_{T})\rho+f_{T}P]
\label{26}
\een
and
\ben
&&2\dot{H}+3H^2=-P_{\rm eff}\n &&=\frac{f}{4F}-\frac{2\dot{F}H}{F}+\frac{1}{2F}[(1+f_{T})\rho+(2+f_{T})P]\n
\label{27}
\een
Thus, the effective thermodynamic quantities adhere to the conventional conservation equation 
\ben
\dot{\rho}_{\rm eff} + 3H\Big(\rho_{\rm eff} + P_{\rm eff}\Big)=0. 
\label{28}
\een

To characterize the cosmic dynamics of the model, it is beneficial to incorporate an effective equation of state (EoS) parameter. The effective equation of state (EoS) parameter is expressed as follows: 
\ben 
w_{\text{eff}} \equiv \frac{P_{\text{eff}}}{\rho_{\text{eff}}} = 1+\frac{2(4\dot{F}H-2P-f)} {f-2[(1+f_T)\rho + f_T P]}. 
\label{29} 
\een
The above expression \eqref{29} denotes the complete effective EoS parameter, integrating contributions from conventional matter, the DBI scalar field, and the geometric alterations stemming from the $f(Q, T)$ sector via Eq. \eqref{22}.\\

An important quantity characterizing the dynamical evolution of the universe is the deceleration parameter $q$, which determines whether the cosmic expansion is accelerating or decelerating. It is defined as
\ben
q \equiv -\frac{\dot{H}}{H^2} - 1 = \frac{1}{2}\big(1 + 3\,\omega_{\rm eff}\big).
\label{30}
\een
A negative value of $q$ corresponds to an accelerating universe, whereas a positive value indicates a decelerating phase of expansion.

In the present $f(Q,T)$ framework \eqref{1}, the deceleration parameter can be expressed explicitly as
\ben
q = -1 + \frac{3\big(4\dot{F}H - 2P - f + 12F H^2\big)}
{f - 2\big[(1+f_T)\rho + f_T P\big]}.
\label{31}
\een\\

This decomposition highlights that the resulting cosmic dynamics differs from both standard cosmology and previously studied $f(Q,T)$ scenarios with simple matter sources. In general relativity, the expansion history is controlled solely by the total energy density, whereas in $f(Q,T)$ gravity the matter--geometry coupling modifies the effective gravitational dynamics and typically induces non-conservation of the matter energy--momentum tensor. The addition of a DBI-type (non-canonical) scalar field introduces an extra dynamical sector with a relativistic kinetic structure, leading to a coupled interplay between the non-metricity sector and scalar-field dynamics in the Friedmann equations. As a consequence, the effective role of pressure in driving (or opposing) cosmic acceleration can be enhanced or suppressed relative to either conventional $f(Q,T)$ gravity or pure DBI cosmology. Moreover, the presence of the $f_T$ coupling can be interpreted as a generalized channel for energy exchange among the matter, geometry, and DBI sectors, opening new possibilities for describing late-time acceleration and unified cosmic evolution within a single framework.

\section{Cosmological Model, Datasets, and Parameter Inference}
\label{S3}
We investigate a cosmological scenario in $f(Q,T)$ gravity and adopt the minimal linear form $f(Q,T)=\alpha Q+\beta T$, where $\alpha=-1$ is a constant and $\beta$ is a dimensionless matter-geometry coupling parameter. In the limit $\beta\to 0$, the model reduces to the Symmetric Teleparallel Equivalent of General Relativity (STEGR), whereas $\beta\neq 0$ yields an explicit interaction between matter and geometry through the trace $T$. Owing to its simplicity and transparent physical meaning, this ansatz has been widely used to explore cosmological dynamics, dark-energy phenomenology, and departures from standard general relativity in non-metricity-based theories of gravity \cite{Xu2019, Harko2018}. 

The cosmic medium is assumed to comprise baryonic perfect-fluid matter and a Dirac-Born-Infeld (DBI) scalar field $\phi$ \eqref{1}; the non-canonical kinetic structure introduces a Lorentz-type factor that can strongly restrict the field evolution. Following standard DBI constructions, we choose a specific warp factor and potential
\ben
 f(\phi)=\lambda \phi^4, \qquad V(\phi)=\frac{1}{2}m^{2}\phi^{2},
 \label{32}
\een
where $\lambda$ and $m$ set the warping strength and the mass scale, respectively \cite{Gumjudpai, Silverstein, Alishahiha, Ganguly}.

The particular form of $f(\phi)$ \eqref{32} ensures the correct DBI dynamics ; it enhances relativistic DBI effects at small $\phi$. The quadratic potential provides a simple massive scalar with a stable minimum at $\phi=0$, and it is widely used due to its analytic transparency and compatibility with slow-roll-type regimes. Together, these ingredients yield a well-motivated and phenomenologically viable setting for exploring cosmological dynamics in modified gravity.\\

We constrain the free parameters of model~\eqref{1} using measurements of the Hubble expansion rate $H(z)$ (cosmic chronometers) \cite{zhang2014four,simon2005constraints,moresco2012improved,moresco20166,stern2010cosmic,moresco2015raising,Ratsimbazafy,Borghi}, DESI BAO observations \cite{Hussain}, and the Pantheon+SH0ES Type~Ia supernova compilation \cite{Brout,Scolnic1}. In the joint analysis, we take the parameter vector to be
\ben
\Theta \equiv (H_0,\Omega_{m_0},\beta,\lambda,\phi_0,x_0,r_d),
\label{33}
\een
where $H_0$ is the present-day Hubble constant, $\Omega_{m_0}$ is the current matter density parameter, $\beta$ and $\lambda$ are the model coupling parameters, $\phi_0\equiv\phi(z=0)$ is the present-day value of the scalar field, and $r_d$ is the comoving sound horizon at the drag epoch. Parameter inference is carried out by maximizing (and sampling) the combined likelihood, assuming Gaussian measurement errors and treating the datasets as statistically independent. We report the corresponding best-fit values together with credible/confidence intervals.

Here, we define the dimensionless variable $x=\frac{\dot{\phi}}{\sqrt{\lambda}\,\phi^2}$, where $\dot{\phi}\equiv\frac{d\phi}{dt}$, which recasts the scalar-field evolution into a compact first-order system. Using the specified forms of $f(\phi)$ and $V(\phi)$ (cf.~Eq.~\eqref{32}), Eq.~\eqref{13} can be written in terms of $x$ as
\ben
\frac{dx}{dz}&&=\frac{1}{H(1+z)}\Bigg[\frac{3Hx}{\gamma^2}+4\sqrt{\lambda}\,\phi\Bigg(1-\frac{1}{\gamma^3}-x^2\Bigg)\n &&+\frac{m^2}{\sqrt{\lambda}\,\phi\,\gamma^3}\Bigg]~,
\label{33a}
\een
where $\gamma\equiv\frac{1}{\sqrt{1-x^2}}$ (cf.~Eq.~\eqref{12}), which is required for our observational data analysis.

\subsection{Markov Chain Monte Carlo (MCMC) Analysis}

We estimate the model parameters by fitting to the observational datasets using MCMC sampling implemented in \texttt{Cobaya} \cite{ref1,ref2,ref3,ref4,ref5,ref6,ref7,ref8,Thakre}. The inference is based on the total likelihood, assumed Gaussian and written in terms of the total chi-square \cite{Ntampaka1},
\ben
\mathcal{L}_{\mathrm{tot}} \propto \exp\left(-\frac{1}{2}\chi^2_{\mathrm{tot}}\right),
\label{34}
\een
where $\chi^2_{\mathrm{tot}}$ denotes the combined contribution of all datasets. The resulting chains sample the posterior $\Theta=(H_0,\Omega_{m0},\beta,m)$ and are used to report best-fit values, confidence intervals, and model-selection diagnostics including $\chi^2$, AIC \cite{Akaike}, BIC \cite{Schwarz}, DIC \cite{spiegelhalter2002} and the relative criteria $\Delta\mathrm{AIC}$, $\Delta\mathrm{BIC}$ and $\Delta\mathrm{DIC}$.

The resulting triangular (corner) plots summarize the parameter constraints: the diagonal panels show the one-dimensional marginalized posteriors, while the off-diagonal panels display the two-dimensional confidence regions, with the inner and outer contours corresponding to $1\sigma$ and $2\sigma$, respectively. Overall, this methodology provides a robust framework for parameter estimation and for assessing the consistency of the theoretical model with the observational data.

\subsection{Hubble Parameter Measurements}
\label{subsec:Hz}

We use 32 cosmic-chronometer measurements of the Hubble expansion rate (Table~\eqref{tab:1}), obtained with the differential-age (DA) method, spanning the redshift interval $0.07 \le z \le 1.965$. This range is well suited to probing the late-time expansion history, as it covers the transition from the recent accelerated epoch to intermediate redshifts where dark-energy effects become progressively important. Cosmic-chronometer observations provide direct, largely model-independent estimates of the Hubble parameter $H(z)$ and are therefore particularly valuable for constraining cosmological dynamics and testing modified-gravity scenarios at low and intermediate redshifts. The expansion rate is defined as \cite{Hogg1,Hogg2}
\ben
H(z)=-\frac{1}{1+z}\frac{dz}{dt},
\label{35}
\een
and the corresponding contribution to the total chi-square is \cite{Hogg1,Hogg2}
\ben
\chi^2_{H}=\sum_{i=1}^{32}\left[\frac{H_{\mathrm{th}}(z_i;\Theta)-H_{\mathrm{obs}}(z_i)}{\sigma_{i}}\right]^2,
\label{36}
\een
where $H_{\mathrm{obs}}(z_i)$ and $\sigma_i$ denote the observed values and their $1\sigma$ uncertainties, respectively, and $H_{\mathrm{th}}(z_i;\Theta)$ is the theoretical prediction. The parameter vector is
\ben
\Theta\equiv\left(H_0,\Omega_{m0},\beta,\lambda, \phi_0,x_0\right).
\label{37}
\een

In this work, $H(z)$ is obtained by numerically integrating the coupled background equations for a DBI scalar field in $f(Q,T)$ gravity. The field variable $\phi$ and its velocity are evolved together with the modified gravitational equations, and the resulting solutions are used to reconstruct the expansion rate at the redshifts of interest.

Unlike in $\Lambda$CDM, where $H(z)$ can be written in closed form, the present setup yields a non-linear dynamical system driven by the non-canonical DBI kinetic term and the explicit matter--geometry coupling in the $f(Q,T)$ sector. As a result, $H(z)$ receives contributions from the baryonic component, the DBI field, and the geometric corrections induced by modified gravity.

The cosmic-chronometer compilation therefore provides a key constraint on the background evolution, offering direct and nearly model-independent measurements of $H(z)$ over a broad redshift range. The observational data used here are reported in Table~\eqref{tab:1}.

\subsection{DESIBAO measurements from DR2}
We use baryon acoustic oscillation (BAO) measurements from the Dark Energy Spectroscopic Instrument (DESI) collaboration \cite{Abdul}. In particular, we adopt the DESI BAO dataset \cite{Hussain} from the most recent DESI data release (DR2; see Table~\eqref{tab:placeholder13}), ensuring that our analysis is based on the latest available observational constraints. In modern cosmology, BAO provides a powerful probe of the expansion history of the universe \cite{5}.

To interpret BAO signatures in large-scale galaxy surveys, several cosmological distance measures are commonly employed, including the Hubble distance $D_H$, the volume-averaged distance $D_V$, and the transverse comoving distance $D_M$. These quantities map the observed angular and redshift separations on the sky to physical length scales, enabling constraints on the Hubble constant $H_0$ and the dark-energy equation of state $\omega(z)$.

The radial BAO information is commonly expressed through the Hubble distance at redshift $z$,
\ben
D_H(z)=\frac{c}{H(z)},
\label{38}
\een
where $H(z)$ is the Hubble parameter and $c$ is the speed of light.

In the transverse direction, the BAO feature subtends an angle $\Delta\theta$ on the sky. This angular separation is related to the comoving sound horizon at the drag epoch, $r_d$, via $r_d=D_M(z)\,\Delta\theta$, where the transverse comoving distance is \cite{5,Percival,Blake}
\ben
D_M(z)=c(1+z)\int_0^z\frac{dz'}{H(z')}.
\label{39}
\een

The quantity $r_d$ provides the standard ruler for BAO analyses and is defined as \cite{Hu2,Eisenstein2}
\ben
r_d = \int_{z_{\rm drag}}^{\infty} \frac{c_s(z)}{H(z)}\,dz,
\label{40}
\een
where $z_{\rm drag}$ is the baryon drag epoch, i.e., the redshift at which baryons kinetically decouple from the photon-baryon plasma, and $c_s(z)$ is the the baryon-photon sound speed.

BAO constraints are often summarized using the spherically averaged (isotropic) distance scale
\ben
D_V(z)=\big[z\,D_M^2(z)\,D_H(z)\big]^{1/3}.
\label{41}
\een

For the BAO dataset, we evaluate the goodness of fit using \cite{Hogg1,Hogg2}
\ben
\chi^2_{\rm BAO}=\sum_{k=1}^N\left(\frac{X_k^{\rm th}(z_k,\Theta)-X_k^{\rm obs}(z_k)}{\sigma_k}\right)^2,
\label{42}
\een
where $X_k^{obs}$ denotes the measured BAO observable (typically $D_M/r_d$, $D_H/r_d$, or $D_V/r_d$) and $\Theta$ is the set of model parameters. The total contribution from BAO data is then taken as
\ben
\chi^2_{\rm BAO}=\chi^2_{(D_M/r_d)}+\chi^2_{(D_V/r_d)}+\chi^2_{(D_H/r_d)}.
\label{43}
\een

\subsection{Pantheon$+$SH0ES Data}

The Pantheon$+$SH0ES compilation provides 1701 light curves for 1550 distinct, spectroscopically confirmed Type~Ia supernovae (SNe~Ia), covering the redshift range $0.00122\le z\le 2.2613$ \cite{Brout,Scolnic1}. However, to omit the peculiar velocity basing, we use only those data ranging from $z=0.01$ to $z=2.2613$\cite{Ganguly}. This includes 1589 data points (SNe Ia). 

We constrain the model parameters by comparing the observed and predicted distance moduli,
\ben
\mu(z,\theta)=m-M=5\log_{10}(d_l(z))+25,
\label{44}
\een
where $d_l(z)$ is the luminosity distance (in Mpc),
\ben
d_l(z)=(1+z)c\int^z_0\frac{dz'}{H(z')}.
\label{45}
\een
Here $m$ is the apparent magnitude and $M$ is the absolute magnitude (defined at 10~pc).

For Pantheon$+$SH0ES, we evaluate the supernova likelihood using the full covariance-matrix approach \cite{Brout,Scolnic1}. After imposing the redshift cut, we extract the corresponding $N_{\rm SN}\times N_{\rm SN}$ submatrix from the released covariance matrix. The resulting chi-square is
\ben
\chi^2_{\rm SN}=\big(\boldsymbol{\mu}_{\rm th}-\boldsymbol{\mu}_{\rm obs}\big)^{\!T}\,\mathbf{C}^{-1}\,\big(\boldsymbol{\mu}_{\rm th}-\boldsymbol{\mu}_{\rm obs}\big),
\label{46}
\een
where $\boldsymbol{\mu}_{\rm th}$ and $\boldsymbol{\mu}_{\rm obs}$ are vectors of theoretical and observed distance moduli for the selected supernovae, respectively. The theoretical prediction $\mu_{\rm th}(z_i,\Theta)$ is computed from Eq.~\eqref{44}.

\section{Result and Discussion}\label{S4}
In this section, we statistically confront the proposed $f(Q,T)$ model with the DBI-type action \eqref{1} against the standard $\Lambda$CDM and $\omega$CDM scenarios using a combined dataset of Hubble measurements, Pantheon+SHOES supernovae, and DESI BAO data. Our aim is to quantify how well each framework reproduces the observations while accounting for the practical impact of increased model flexibility on the fit quality and robustness, as well as the resulting tightness of the parameter constraints in our model. The main statistical indicators are reported in Table \eqref{tab:placeholder}, and the corresponding best-fit cosmological parameters are provided in Table \eqref{tab:placeholder11}.\\

\begin{figure}[H]
\centering
\includegraphics[width=7.9cm, height=7.8cm]{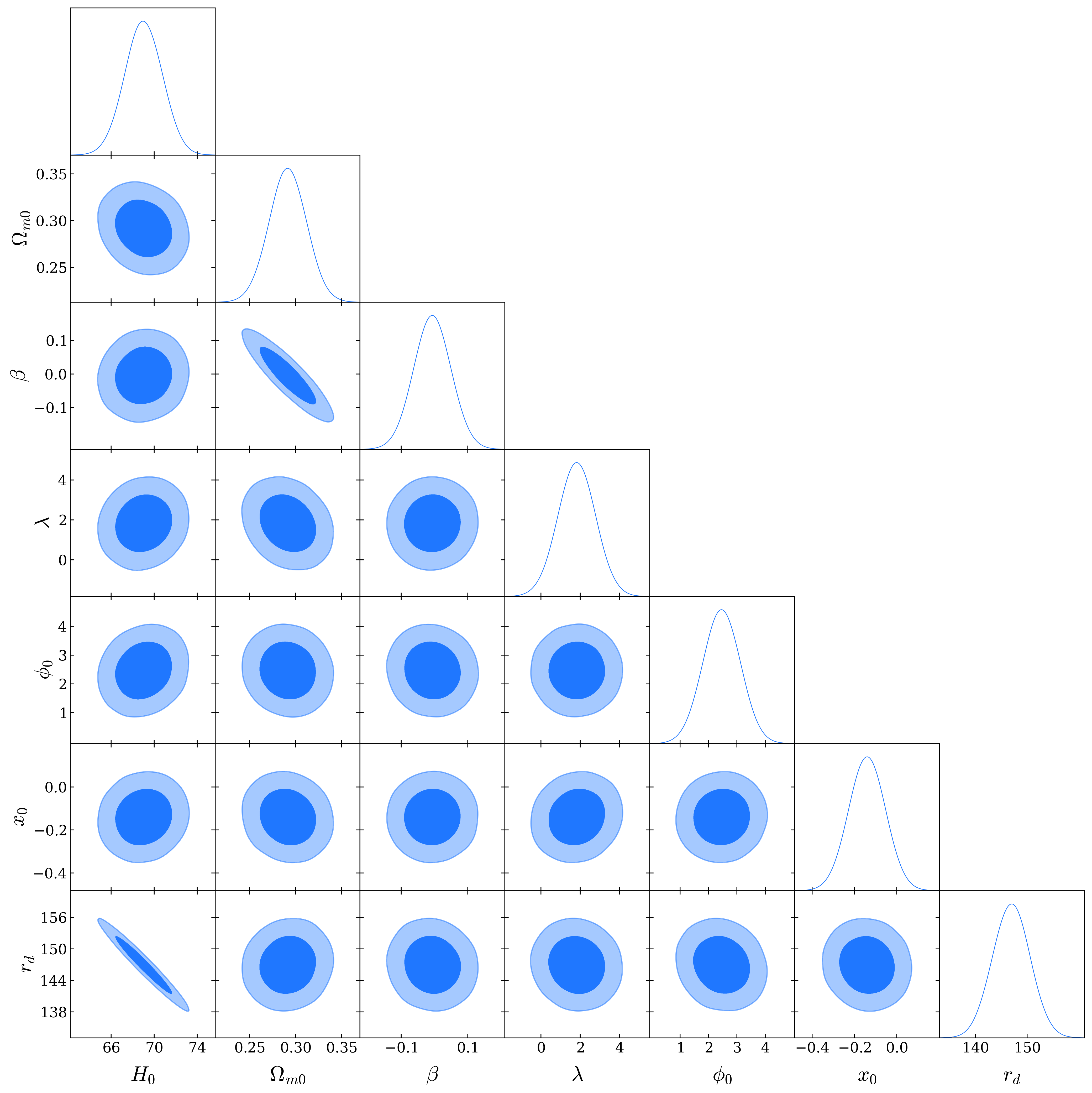}
\caption{Bayesian inference using a Gaussian approximation of the model parameters for the combined data set Hubble, Pantheon+SH0ES, and DESIBAO for the $f(Q,T)+$ DBI Model.}
\label{HpdbT}
\end{figure}

\begin{table*}[t]
    \centering
    \renewcommand{\arraystretch}{1.4}
\begin{tabular}{c c c }\hline \hline
\textbf{Parameter}  & \textbf{Prior} & \textbf{Best Fit} \\ \hline 
$H_0[Km/s/Mpc]$&  $\mathcal{U}$(60,80)& 69.231$\pm$1.738\\
$\Omega_{m_0}$&  $\mathcal{U}$(0.2,0.4)& 0.319$\pm$0.020\\
 $r_d$& $\mathcal{U}$(130,170)&146.253$\pm$3.615\\ \hline
 $\beta$&  $\mathcal{U}$(-0.1,0.1)& -0.070$\pm$0.057\\
 $\lambda$& $\mathcal{U}(0.5,4.0)$&0.548$\pm$0.951\\
 $\phi_0$& $\mathcal{U}(1.5,4.5)$&2.134$\pm$0.661\\
 $x_0$& $\mathcal{U}(-0.30,0.0)$&-0.216$\pm$0.086\\ \hline \hline
\end{tabular}
\caption{Parameter estimation and uncertainties using Bayesian inference with the MCMC analysis method for $f(Q,T)+$DBI model.}
\label{tab:placeholder22}
 \end{table*}

Figure~\eqref{HpdbT} shows the corner plot of the marginalized posterior distributions obtained from the MCMC analysis of the $f(Q,T)+\mathrm{DBI}$ model using the combined Hubble, Pantheon+SHOES, and DESI BAO data. The diagonal panels display the one-dimensional posteriors for the full parameter set $(H_0,\Omega_{m_0},\beta,\lambda,\phi_0,x_0,r_d)$, while the off-diagonal panels show the corresponding two-dimensional $1\sigma$ and $2\sigma$ credible regions.

The constraints reported in Table~\ref{tab:placeholder22} can be directly compared with standard late-time cosmological measurements.
The inferred matter density, $\Omega_{m_0}=0.319\pm0.020$, lies in the range typically preferred by low-redshift probes and is consistent with values obtained in $\Lambda$CDM analyses using BAO and SNe, as well as with CMB-inferred estimates once model assumptions are taken into account.
The coupling parameter of the $f(Q,T)$ sector is $\beta=-0.070\pm0.057$, statistically compatible with zero, indicating that the current data do not require a significant matter--geometry coupling and that the model approaches the symmetric-teleparallel/GR-like limit at late times.
In the DBI sector, the parameters $\lambda=0.548\pm0.951$ and $\phi_0=2.134\pm0.661$ are moderately well constrained, showing that the combined Hubble, Pantheon+SHOES, and DESI BAO dataset is sensitive to the scalar-field contribution.
Moreover, $x_0=-0.216\pm0.086$ is concentrated near zero, implying that the field is slowly rolling today and that the late-time dynamics is close to an effective cosmological-constant behavior.
The sound-horizon scale, $r_d=146.253\pm3.615\,\mathrm{Mpc}$, is nearly consistent with values typically inferred from early-universe calibrated $\Lambda$CDM analyses, indicating that the model preserves the standard BAO ruler while allowing deviations from $\Lambda$CDM through the late-time modified-gravity sector.
The relatively tight constraint on $r_d$ also highlights the constraining power of BAO measurements and their complementarity with expansion-history data.

The joint analysis gives $H_0=69.231\pm1.738 ~\rm km ~s^{-1}Mpc^{-1}$ , a value intermediate between the CMB-inferred and local distance-ladder estimates. This behavior can be attributed to the additional degrees of freedom associated with the DBI scalar field and the $f(Q,T)$ coupling, which modify the late-time cosmological dynamics. In particular, BAO constraints are primarily sensitive to combinations such as $D_M(z)/r_d$ and $H(z)\,r_d$; therefore, a modest reduction in $r_d$ can correlate with an increased inference $H_0$ while still providing an acceptable fit to the BAO measurements. Physically, a smaller effective $r_d$ relative to the early-universe $\Lambda$CDM value can be interpreted as mimicking mechanisms that shorten the sound horizon (e.g., an effectively faster pre-recombination expansion or a rescaling of the BAO ruler), although in our analysis it emerges as an effective late-time parameter in the combined fit rather than from an explicit early-time model.

Differences between our best-fit values and those reported in other studies can arise simply because different observational datasets are used. This can reflect (i) differences in sample selection and redshift coverage (late-time-only compilations versus analyses that also include early-universe information such as CMB priors), (ii) cross-calibration choices and the treatment of nuisance parameters (e.g., the Pantheon+SHOES absolute-magnitude calibration, light-curve standardization, and covariance modeling), (iii) differences in BAO methodology (isotropic versus anisotropic measurements, reconstruction choices, and assumptions used to convert observed angles/redshifts into distances), (iv) systematic uncertainties in cosmic chronometers (stellar-population synthesis modeling, metallicity, and chronometer selection), and (v) how correlations between datasets are incorporated in the likelihood (overlapping sky coverage, shared systematics, and potential covariance double-counting). In extended models such as $f(Q,T)+\mathrm{DBI}$, parameter degeneracies—most notably among $(H_0,r_d,\Omega_{m_0})$ and the DBI-sector parameters $(\lambda,\phi_0,x_0)$—can amplify these effects, so that even modest shifts in the preferred distance scale or expansion history in one dataset lead to noticeably different posterior means when compared across dataset combinations.

\begin{figure}[H]
\centering
\includegraphics[width=7.9cm, height=7.8cm]{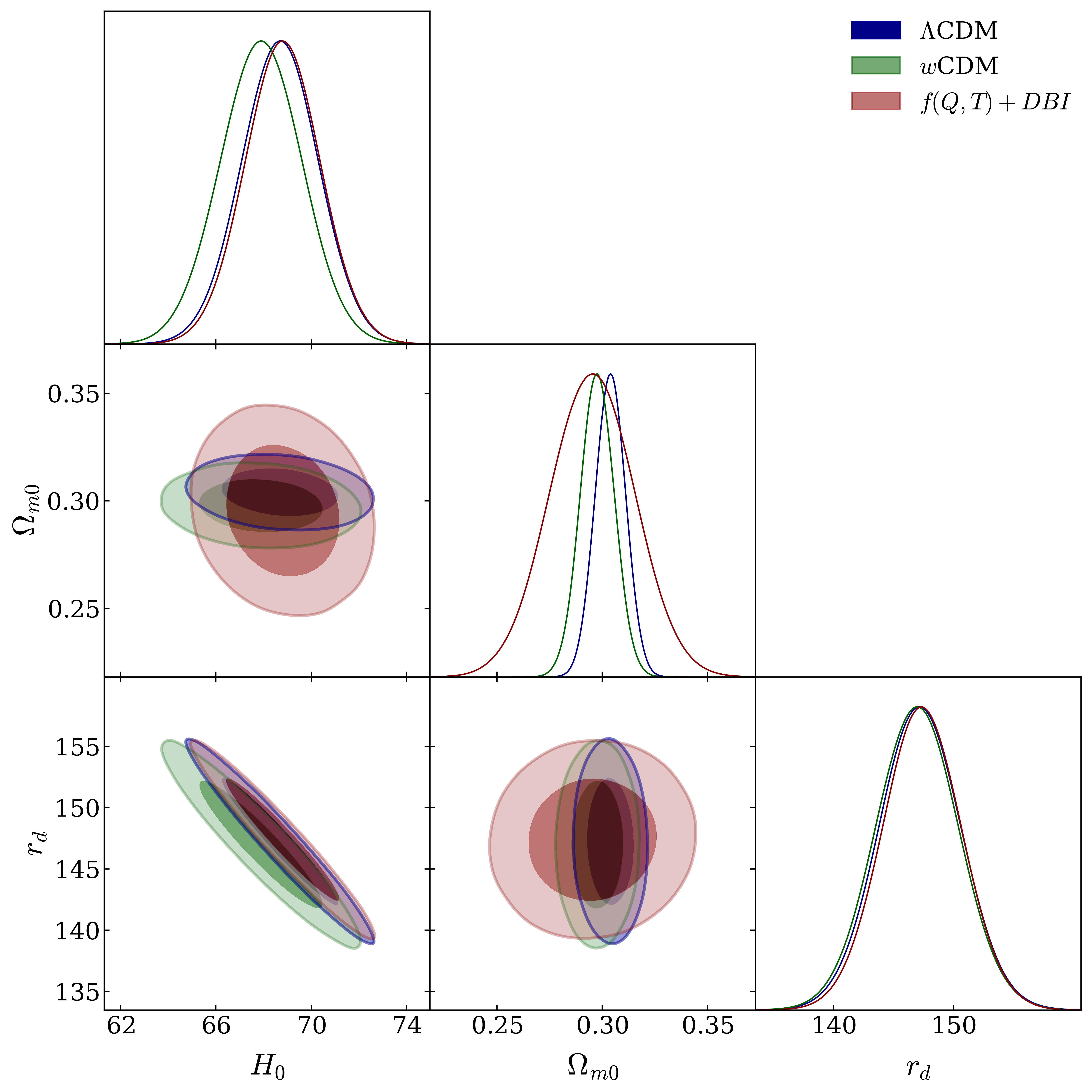}
\caption{Bayesian inference for all models with Gaussian approximation of the model parameters for the combined data set Hubble, Pantheon+SH0ES and DESIBAO.}
\label{mpl}
\end{figure}

\begin{figure}[H]
\centering
\includegraphics[width=7.9cm, height=7.0cm]{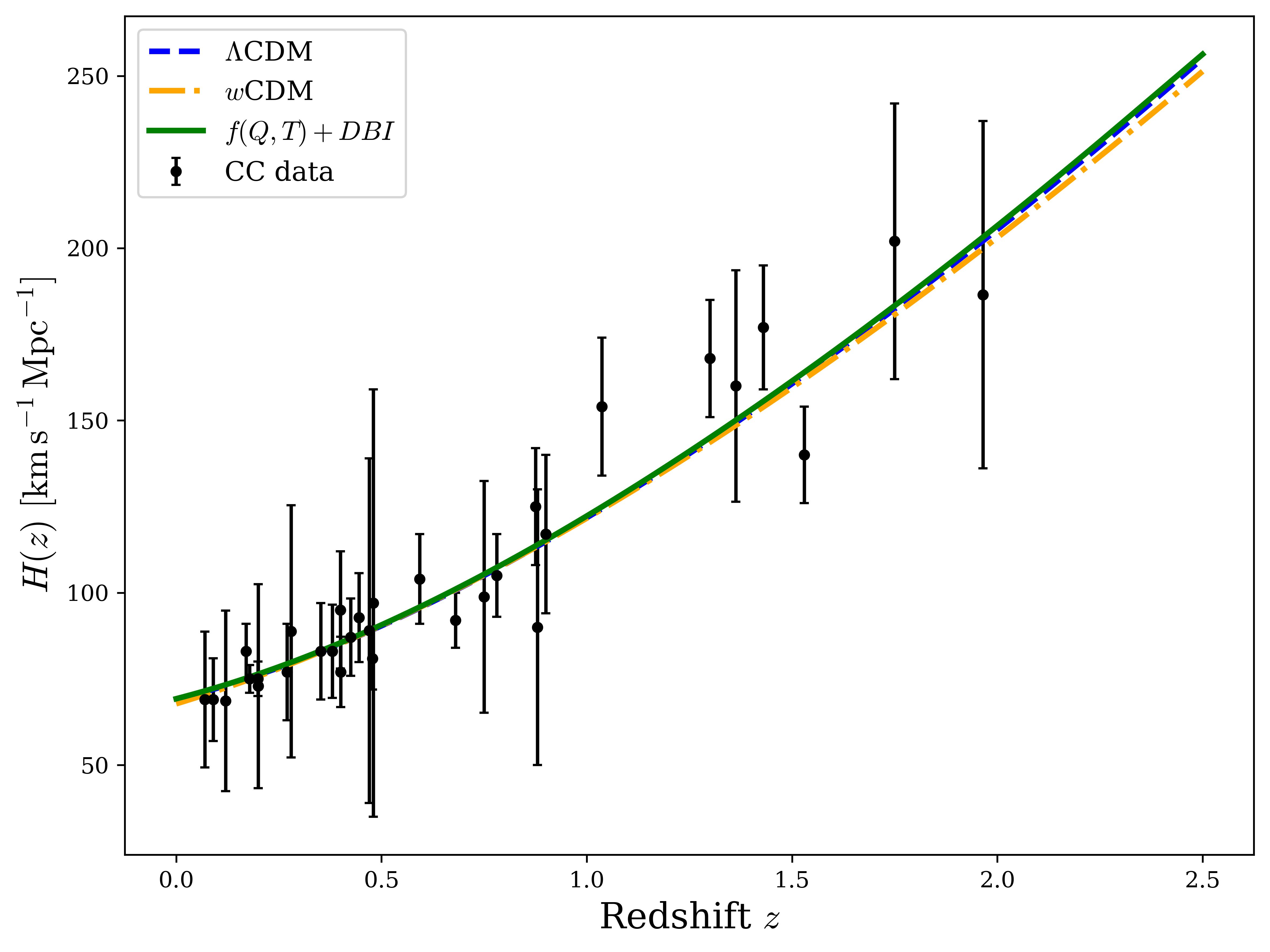}
\caption{Comparison of Hubble cosmic chronometer data fits for all models.}
\label{hzzz}
\end{figure}

\begin{table*}[t]
    \centering
    \renewcommand{\arraystretch}{1.4}
    \begin{tabular}{c c c c}\hline\hline
         Model&  $H_0[Km/s/Mpc]$&  $\Omega_{m_0}$& $r_d$\\\hline
         $\Lambda$CDM&  68.932$\pm$1.867&  0.303$\pm$0.007& 146.735$\pm$3.847\\
         $\omega$CDM&  67.777$\pm$2.269&  0.298$\pm$0.008& 147.184$\pm$4.415\\
         $f(Q,T)+\mathrm{DBI}$&  69.231$\pm$1.738&  0.319$\pm$0.020& 146.253$\pm$3.615\\\hline
    \end{tabular}
    \caption{Comparative table of cosmological Parameters for Different Models}
    \label{tab:placeholder11}
\end{table*}

Table \ref{tab:placeholder11} lists the best-fit constraints on the Hubble constant $H_0$, the present-day matter density $\Omega_{m_0}$, and the sound-horizon scale $r_d$ for $\Lambda$CDM, $w$CDM, and our $f(Q,T)+\mathrm{DBI}$ model, using the combined Hubble, Pantheon+SH0ES, and DESIBAO datasets. For the standard scenarios, we obtain $H_0\simeq 68.932\pm1.867\,\mathrm{km\,s^{-1}\,Mpc^{-1}}$, $\Omega_{m_0}\simeq 0.303\pm0.007$, and $r_d\simeq 146.735\pm3.847\,\mathrm{Mpc}$ in $\Lambda$CDM, and $H_0\simeq 67.777\pm2.269\,\mathrm{km\,s^{-1}\,Mpc^{-1}}$, $\Omega_{m_0}\simeq 0.298\pm0.008$, and $r_d\simeq 147.184\pm4.415\,\mathrm{Mpc}$ in $w$CDM.

Figure \eqref{mpl} shows a comparative Bayesian analysis of the $\Lambda$CDM, $w$CDM, and $f(Q,T)+\mathrm{DBI}$ cosmological models in the shared parameter space $(H_0,\Omega_{m_0},r_d)$. The posterior confidence contours largely overlap at the $1\sigma$ and $2\sigma$ levels, indicating that all three scenarios fit the current data comparably well. While the $f(Q,T)+\mathrm{DBI}$ model exhibits a mild preference for larger $\Omega_{m_0}$, the substantial overlap of the credible regions confirms that it is statistically consistent with both $\Lambda$CDM and $w$CDM.

Figure \eqref{hzzz} compares the Hubble parameter evolution $H(z)$ predicted by the $\Lambda$CDM, $w$CDM, and $f(Q,T)+\mathrm{DBI}$ models with cosmic chronometer measurements. All three theoretical curves track the data across the full redshift range, with only minor differences between the models. This agreement shows that the proposed $f(Q,T)+\mathrm{DBI}$ scenario reproduces the observed expansion history and remains competitive with the standard cosmological descriptions.\\

The statistical parameters used in the analysis are defined as follows:
\begin{itemize}
    \item $\chi^2$ (chi-square statistic): Quantifies the mismatch between the observational data and the theoretical predictions. Smaller values indicate a better fit.

    \item $\chi^2_{\rm red}$ (reduced chi-square): Defined as $\chi^2_{\rm red}=\frac{\chi^2}{N-k}$, where $N$ is the total number of data points and $k$ is the number of free model parameters. Values close to unity typically indicate an acceptable fit.

    \item AIC (Akaike Information Criterion): Defined as $\mathrm{AIC}=\chi^2+2k$.

    \item BIC (Bayesian Information Criterion): Defined as $\mathrm{BIC}=\chi^2+k\ln N$.

    \item DIC (Deviance Information Criterion): Defined as $\mathrm{DIC}=\bar{D}(\theta)+p_D$, where $\bar{D}(\theta)$ is the posterior mean deviance and $p_D$ is the effective number of parameters.

    \item $\Delta\mathrm{AIC}$, $\Delta\mathrm{BIC}$, and $\Delta\mathrm{DIC}$: Differences in AIC, BIC, and DIC relative to the reference $\Lambda$CDM model.
\end{itemize}

\begin{table*}[t]
    \centering
    \renewcommand{\arraystretch}{1.4}
    \begin{tabular}{c c c c c c c c c}\hline\hline
            \textbf{Model} &  $\chi^2$ &  $\chi^2_{red}$ &  \textbf{AIC} & \textbf{BIC} & \textbf{DIC} &  $\Delta$\textbf{AIC} &  $\Delta$ \textbf{BIC} & $\Delta$ \textbf{DIC} \\\hline
         $\Lambda$CDM&  1449.04&  0.886&  1457.04&  1478.65&  1457.18&  0&  0& 0\\
         $\omega$CDM&  1445.40&  0.884&  1455.40&  1482.41&  1455.55&  -1.64&  3.76& -1.63\\
         $f(Q,T)+\mathrm{DBI}$&  1449.44&  0.888&  1465.44&  1508.66&  1461.05&  8.39&  30.0& 3.87\\ \hline
    \end{tabular}
    \caption{Comparison of the statistical criteria for the $f(Q,T) + $DBI  Model with $\Lambda$CDM and $\omega$CDM using the MCMC-based Bayesian inference method}
    \label{tab:placeholder}
\end{table*}

In Table \eqref{tab:placeholder}, the likelihood is built from a joint compilation of $N=1640(\rm CC=32, DESI BAO=19, Pantheon+SHOES=1589)$ measurements drawn from four complementary probes: (i) the Hubble $H(z)$ sample constrains the expansion rate directly at multiple redshifts; (ii) the Pantheon Type-Ia supernovae provide distance moduli (relative luminosity distances) over a broad redshift range; (iii) the SH0ES data anchor the local distance ladder through a direct determination of $H_0$; and (iv) the DESI BAO measurements constrain the comoving/angular-diameter distance and the radial distance scale (via the BAO feature) as functions of redshift. Fitting these datasets simultaneously breaks parameter degeneracies by combining absolute and relative distance information with direct expansion-rate constraints. The number of free parameters used in the information criteria is $k=7$ for the $f(Q,T)$+DBI model, $k=3$ for $\Lambda$CDM, and $k=4$ for $\omega$CDM.

To evaluate and compare the performance of the proposed model, we consider the standard goodness-of-fit statistic $\chi^2$, the reduced chi-square $\chi^2_{\rm red}$, and the information criteria $\rm AIC$, $\rm BIC$ and $\rm DIC$, together with their relative differences $\Delta\rm AIC$, $\Delta\rm BIC$ and $\Delta \rm DIC$. These quantities assess the fit quality while penalizing unnecessary model complexity, providing a balanced basis for model selection.\\

Table \eqref{tab:placeholder} summarizes the MCMC-based statistical comparison between $\Lambda$CDM, $w$CDM, and the proposed $f(Q,T)+\mathrm{DBI}$ model. Since $\Lambda$CDM yields the smallest information-criterion values, we adopt it as the reference and set $\Delta\rm AIC=\Delta\rm BIC=\Delta\rm DIC=0$. Relative to this baseline, $w$CDM offers only a marginal improvement in the deviance-based criteria ($\Delta\rm AIC=-1.64$ and $\Delta\rm DIC=-1.63$), but the additional free parameter is penalized by BIC, which increases to $\Delta\rm BIC=3.76$, indicating weak evidence in favor of $w$CDM. The proposed $f(Q,T)+\mathrm{DBI}$ scenario, which enlarges the parameter space through the non-metricity-matter coupling and the generalized DBI scalar-field sector, incurs a stronger complexity penalty; correspondingly, with respect to $\Lambda$CDM we obtain $\Delta\mathrm{AIC}=8.39$, $\Delta\mathrm{BIC}=30$, and $\Delta\mathrm{DIC}=3.87$.

Within the usual AIC heuristic, $\Delta\mathrm{AIC}<2$ indicates statistical indistinguishability, $4<\Delta\mathrm{AIC}<7$ corresponds to considerably weaker support, and $\Delta\mathrm{AIC}>10$ is commonly interpreted as strong evidence against the model \cite{burnham2002}. Physically, AIC asks whether the data require additional degrees of freedom beyond the minimal late-time expansion history described by $\Lambda$CDM: a more flexible model can always lower $\chi^2$ slightly by absorbing small fluctuations and residual systematics, but AIC penalizes that flexibility to favor models that generalize rather than simply fit noise. The value $\Delta\mathrm{AIC}=8.39$ therefore places the $f(Q,T)+\mathrm{DBI}$ model in the ``moderately disfavored'' regime relative to $\Lambda$CDM, meaning that the improvement in fit achieved by the extra modified-gravity and scalar-field parameters is not large enough to justify the added freedom.

An even stronger preference for simplicity is obtained from the Bayesian Information Criterion, for which we find $\Delta\mathrm{BIC}=30$. In Bayesian terms, BIC approximates the competition between the increased likelihood and the ``Occam penalty'' associated with spreading prior weight over a larger parameter space: when many parameters are introduced, only a sufficiently sharp and substantial improvement in the fit can overcome this penalty. Because the BIC penalty grows with both the number of parameters and the size of the dataset, it is particularly restrictive for extended cosmologies confronted with large joint compilations. According to the Kass--Raftery scale, $\Delta\mathrm{BIC}>10$ constitutes very strong evidence in favour of the simpler alternative \cite{kass1995}; hence the combined Hubble, Pantheon+SH0ES, and DESI BAO data strongly prefer the $\Lambda$CDM baseline over the proposed $f(Q,T)+\mathrm{DBI}$ scenario.

The $w$CDM extension illustrates the same trade-off in a milder form: allowing the dark-energy equation of state to deviate from $w=-1$ slightly improves the raw goodness of fit (with somewhat smaller $\chi^2$, AIC, and DIC than $\Lambda$CDM), but the improvement is too small to compensate for the additional parameter once model complexity is accounted for, so BIC increases. Overall, the data provide only weak support for $w$CDM over $\Lambda$CDM. 

By contrast, the DIC leads to a more nuanced conclusion for the modified-gravity scenario. The value $\Delta\mathrm{DIC}=3.87$ indicates only weak-to-moderate support for the model with the smaller DIC, rather than a decisive preference; in this sense, the predictive performance of the proposed $f(Q,T)+\mathrm{DBI}$ model remains broadly comparable to that of $\Lambda$CDM. Unlike AIC and BIC, which are largely driven by the best-fit likelihood and the nominal parameter count, DIC is computed from the full posterior and penalizes the effective number of parameters actually constrained by the data \cite{spiegelhalter2002}. As such, it offers a complementary Bayesian perspective, particularly relevant for models with extended parameter spaces and strong parameter degeneracies.

This behavior can be understood from the structure of the model itself. Although the $f(Q,T)+\mathrm{DBI}$ framework introduces several additional parameters ($k=7$ versus $k=3$ in $\Lambda$CDM; Table~\eqref{tab:placeholder}), the posterior constraints suggest that not all of these directions in parameter space are effectively constrained by the current data combination. As a result, the model's effective complexity can be smaller than its nominal parameter count, so the DIC penalty (which depends on the effective number of parameters) grows more mildly than the AIC/BIC penalties. In this sense, the comparatively large values $\Delta\mathrm{AIC}=8.39$ and $\Delta\mathrm{BIC}=30$ mainly reflect the increased formal complexity of the extended framework, whereas the more moderate $\Delta\mathrm{DIC}=3.87$ indicates that its posterior-averaged predictive performance remains broadly competitive with the standard $\Lambda$CDM cosmology.

Moreover, all three models deliver comparably good fits, as reflected by reduced chi-square values close to unity: $\chi^2_{\mathrm{red}}=0.886$ for $\Lambda$CDM, $\chi^2_{\mathrm{red}}=0.884$ for $w$CDM, and $\chi^2_{\mathrm{red}}=0.888$ for the proposed $f(Q,T)+\mathrm{DBI}$ scenario. This indicates that each framework can reproduce the combined late-time data at an acceptable statistical level. The stronger AIC and BIC penalties incurred by $f(Q,T)+\mathrm{DBI}$ should therefore be viewed mainly as a consequence of its enlarged parameter space, rather than as a sign of a poor fit.

Overall, although the information criteria favor the simpler benchmark cosmologies, the $f(Q,T)+\mathrm{DBI}$ model remains observationally viable. In particular, the moderate value $\Delta\mathrm{DIC}=3.87$ suggests that its posterior-averaged predictive performance is broadly comparable to that of $\Lambda$CDM, while providing a richer theoretical setting through the nonmetricity-matter coupling and the generalized DBI scalar-field sector.\\

We summarize our results as follows. Table~\ref{tab:1} lists the $H(z)$ compilation used in our analysis. Table~\ref{tab:placeholder} reports the goodness-of-fit statistics and information criteria for the $f(Q,T)+\mathrm{DBI}$ model using the combined $H(z)$, Pantheon+SHOES, and DESIBAO datasets, while Table~\ref{tab:placeholder11} provides the corresponding best-fit cosmological parameters for the different scenarios considered. Table~\ref{tab:placeholder13} summarizes the DESIBAO measurements from DESI Data Release~2 (DR2). Additional results for alternative dataset combinations are presented in Appendix~\ref{appen:C}. Finally, Table~\ref{stat} collects the best-fit parameters and statistical indicators for several subsets of the data, namely $H(z)$, $H(z)+\mathrm{BAO}$, $H(z)+\mathrm{Pantheon}$, and $H(z)+\mathrm{DESIBAO}$. As expected, combining complementary probes tightens the parameter constraints; in particular, the inferred values of $H_0$ and $\Omega_{m0}$ remain within the ranges typically found in standard cosmological analyses.

\section{Conclusion}\label{S5}
In this work, we investigated an accelerating cosmological scenario in the framework of $f(Q,T)+\mathrm{DBI}$ gravity. In $f(Q,T)$ theory, a non-minimal coupling between geometry and matter can arise, leading to a non-conserved energy-momentum tensor. This coupling introduces additional gravitational effects that may impact the late-time dynamics of the universe and could contribute to the observed cosmic acceleration. We focused on the linear model $f(Q,T)= \alpha Q+\beta T$, where $Q$ is the non-metricity scalar and $T$ is the trace of the energy-momentum tensor. The cosmological evolution has been examined in a spatially flat FLRW background, with the DBI-essence scalar field playing the role of an effective dark-energy component.

After deriving the corresponding cosmological field equations, we constrained the model parameters using Bayesian MCMC techniques. Observational constraints were obtained from Hubble measurements, the Pantheon+SHOES Type Ia supernova compilation, and DESI BAO (DR2) data. The inferred parameter space has been examined via posterior distributions, confidence contours, triangle plots, and uncertainty estimates. We also performed a statistical comparison with the standard $\Lambda \rm CDM$ and $\omega \rm CDM$ models using the $\chi^2$, $\chi^2_{\rm red}$, AIC, BIC, and DIC criteria.

Our statistical analysis indicates that all three cosmological scenarios fit the data well, with $\chi^2_{\rm red}$ remaining close to unity. Once model complexity is taken into account, $\Lambda \rm CDM$ and $\omega \rm CDM$ are mildly favoured by the information criteria; nevertheless, the proposed model remains observationally viable and statistically competitive. In particular, the DIC value suggests that the modified-gravity scenario has predictive performance very similar to that of the standard cosmological model, even though it introduces additional free parameters. The higher AIC and BIC values for the proposed model therefore mainly reflect the stronger penalty for its larger parameter space, rather than a meaningful deterioration in the quality of the fit.

As summarized in Table~\ref{tab:placeholder}, the inferred parameter bounds are broadly consistent with current cosmological constraints, suggesting that a non-metricity-matter coupling can influence the late-time dynamics in a phenomenologically relevant way. Within this framework, the generalised DBI-essence sector can source the observed accelerated expansion while remaining compatible with the datasets considered here. At the same time, we emphasise that our model is not expected to match every observational feature exactly; rather, it provides an alternative modified-gravity avenue, with additional theoretical freedom beyond $\Lambda$CDM, that remains statistically and observationally viable.

We do not claim that the present scenario offers a complete resolution of all cosmological tensions. Instead, it represents a plausible extension of the standard paradigm that may contribute to alleviating discrepancies such as the $H_0$ tension. A more definitive assessment will require joint analyses including CMB measurements, structure-growth data, weak-lensing observations, and large-scale-structure probes, which can tighten the parameter constraints and further clarify the role of non-metricity-matter coupling in the presence of a DBI scalar field, which is beyond the scope of this present work.\\

\begin{acknowledgments}
PKD and GM acknowledge the Inter-University Centre for Astronomy and Astrophysics (IUCAA), Pune, India, for providing them with a visiting associateship under which part of this work was carried out. SI alone acknowledges support from the Deanship of Scientific Research, Vice Presidency for Graduate Studies and Scientific Research, King Faisal University, Saudi Arabia (Grant No.: KFU263108). GM thank the COST Association (CA21136 CosmoVerse), European Union, for the opportunity to participate in this association as a member and express gratitude to undergraduate, postgraduate, and doctoral students, as well as teachers, collaborators, and well-wishers, for their support. \\

\textbf{Conflicts of interest:} The authors declare no conflicts of interest.\\

\textbf{Funding information:} SI thanks the Deanship of Scientific Research, Vice Presidency for Graduate Studies and Scientific Research, King Faisal University, Saudi Arabia, for financial support (Grant No.: KFU263108). \\

\textbf{Data availability:} The data used in this study are readily accessible from public sources for validation of our model; however, we did not generate any new data sets for this research.\\

\end{acknowledgments}

\appendix

\section{Derivation of modified field equation from the action \texorpdfstring{\eqref{1}}{(1)}}
\label{A}

We consider the action
\ben
S= \int \Big[\frac{1}{2}f(Q,T)+\mathcal{L}_{\rm m}+\mathcal{L}_{\rm DBI}(\phi,X)\Big]\sqrt{-g}~d^4x\n
\label{A1}
\een
Now taking the variation of the above Eq. \eqref{A1}, we get 
\ben
\delta S=\int\left[\frac{1}{2}\delta(f\sqrt{-g})+\delta(\mathcal{L}_{\rm m}\sqrt{-g})+\delta(\mathcal{L}_{\rm DBI}\sqrt{-g})\right]d^4x \n
\label{A2}
\een
The contribution of the first term of the above Eq. \eqref{A2} can be written as
\ben
\frac{1}{2}\delta (f\sqrt{-g})&&= \frac{1}{2}\sqrt{-g}~(f_Q\delta Q+f_T \delta T)\n &&-\frac{1}{4}\sqrt{-g}~f\,g_{\mu\nu}\delta g^{\mu\nu}
\label{A3}
\een 
with $\delta f= f_Q\delta Q+f_T \delta T$ and $\delta\sqrt{-g} =-\frac{1}{2}\sqrt{-g}~g_{\mu\nu}\delta g^{\mu\nu}$, where $f_{Q}=\frac{\partial f}{\partial Q}$ and $f_{T}=\frac{\partial f}{\partial T}$.

Also, by the definition \cite{Xu2019}, we have
\ben
T^{(\rm m)}_{\mu \nu}=-\frac{2}{\sqrt{-g}}\frac{\delta (\sqrt{-g}\mathcal{L}_{\rm m})}{\delta g_{\mu\nu}}~;~\Theta_{\mu\nu}=g^{\alpha\beta}\frac{\delta T_{\alpha\beta}}{\delta g_{\mu\nu}}\n
\label{A4}
\een 
with $\delta T= \delta(T_{\mu\nu}g^{\mu\nu})=(T_{\mu\nu}+\Theta_{\mu\nu})\delta g^{\mu \nu}$. So that $\delta (\mathcal{L}_{\rm m}\sqrt{-g})=-\frac{1}{2}\sqrt{-g}~T^{(\rm m)}_{\mu\nu}\delta g^{\mu\nu}$.

On the other hand, we can define for the DBI scalar field \cite{Pal, Mandal, gm3, gm4, Ganguly, Vikman, Martin}
\ben
T^{\rm DBI}_{\mu \nu}=-\frac{2}{\sqrt{-g}} \frac{\delta (\sqrt{-g}\La_{\rm DBI})}{\delta g^{\mu \nu}}
\label{A5}
\een    
so that $\delta (\sqrt{-g}\La_{\rm DBI})=-\frac{1}{2}\sqrt{-g}~T^{\rm DBI}_{\mu\nu}\delta g^{\mu\nu}$.
 Again, we can write \cite{Xu2019}
\ben
\delta Q= 2P_{\alpha \mu \nu}\nabla^\alpha \delta g^{\mu\nu}-(P_{\mu \alpha \beta }Q^{\alpha\beta}_\nu-2Q^{\alpha\beta}{}_\mu P_{\alpha\beta\nu})\delta g^{\mu \nu}.\n
\label{A6}
\een

Finally, following \cite{Xu2019, Heisenberg, Mandal}, using all the above equations and the variation principle, we get the modified field equation as
\ben
&&\frac{-2}{\sqrt{-g}}\nabla^\alpha (f_Q\sqrt{-g}~P^{\alpha}_{\mu\nu})-\frac{1}{2}f g_{\mu \nu}+f_T(T_{\mu\nu}+\Theta_{\mu\nu})\n &&-f_Q(P_{\mu \alpha \beta }Q^{\alpha\beta}_\nu-2Q^{\alpha\beta}{}_\mu P_{\alpha\beta\nu})= T^{(\rm m)}_{\mu\nu}+T^{\rm DBI}_{\mu\nu}.\n
\label{A7}
\een  

\section{Hubble and BAO data set}
The Hubble parameter measurements $H(z)$ listed in Table~\ref{tab:1} are compiled from multiple observational studies covering the redshift range $z\sim 0.07$ to $z\sim 1.965$. They are obtained with the cosmic chronometer (differential-age) technique, which infers $H(z)$ from the relative age evolution of passively evolving galaxies across redshift.

\begin{table*}
    \centering
    \begin{tabular}{|c|c|c|c|l|l|l|l|l|l|l|l|}\hline
    z&  H(z)&  $\sigma_H$& Ref. & z& H(z)& $\sigma_H$& Ref.  & z& H(z)& $\sigma_H$& Ref.  \\\hline
         0.070&  69&  19.6& \cite{zhang2014four} & 0.400& 95 & 17 &\cite{simon2005constraints} & 0.8754& 125& 17&\cite{moresco2012improved} 
\\\hline
         0.090&  69&  12& \cite{simon2005constraints} & 0.4004& 77 & 10.2 &\cite{moresco20166} &  
0.880& 90& 40&\cite{stern2010cosmic} 
\\\hline
         0.120&  68.6&  26.2& \cite{zhang2014four} & 0.4247& 87.1 & 11.2&\cite{moresco20166} & 0.900& 117& 23&\cite{simon2005constraints} 
\\\hline
         0.170&  83&  8& \cite{simon2005constraints} & 0.4497& 92.8& 12.9&\cite{moresco20166} &  
1.037& 154& 20&\cite{moresco2012improved} 
\\\hline
         0.1791&  75&  4& \cite{moresco2012improved} & 0.3802& 83 & 13.5  &\cite{moresco20166} & 1.300& 168& 17&\cite{simon2005constraints} 
\\\hline
         0.1993&  75&  5& \cite{moresco2012improved} & 0.4783& 80.9& 9&\cite{moresco20166}&  
1.363& 160& 33.6&\cite{moresco2015raising} 
\\\hline
         0.200&  7209&  29.6& \cite{zhang2014four} & 0.480& 97& 62&\cite{stern2010cosmic}& 1.430& 177& 18&\cite{simon2005constraints} 
\\\hline
         0.270&  77&  14& \cite{simon2005constraints} & 0.593& 104& 13&\cite{moresco2012improved}&  
1.530& 140& 14&\cite{simon2005constraints} 
\\\hline
         0.280&  88.8&  36.6& \cite{zhang2014four} & 0.6797& 92& 8&\cite{moresco2012improved}& 1.750& 202& 40&\cite{simon2005constraints} 
\\\hline
 0.3519&83 &14 & \cite{moresco2012improved} & 0.7812& 105& 12&\cite{moresco2012improved} 
&  
 1.965& 186.5& 50.4&\cite{moresco2015raising}  \\\hline
 0.47& 89& 50& \cite{Ratsimbazafy}& 0.75& 98.8& 33.6& \cite{Borghi}& & & &\\\hline
\end{tabular}
\caption{The Hubble parameter is reported in units of $\mathrm{km\,s^{-1}\,Mpc^{-1}}$. The $H(z)$ values are inferred using the cosmic chronometer (differential-age) method, and the data sources are listed in the \textit{Ref} column.}
\label{tab:1}
\end{table*}

In this work, we present a detailed analysis of baryon acoustic oscillation (BAO) measurements. We focus on seven BAO data points reported by the Dark Energy Spectroscopic Instrument (DESI) collaboration \cite{Abdul}, which we collectively denote as DESIBAO. Our analysis uses the DESI Data Release~2 (DR2), so that the inferred constraints reflect the most up-to-date DESI measurements available to this study. We adopt DESIBAO as our fiducial BAO dataset; the individual measurements, along with their redshifts and quoted uncertainties, are summarized in Table~\ref{tab:placeholder13}.

\section{Choice of Parameters for the Model}
For the proposed model, we used the the parameter set $(H_0, \Omega_{m_0},\beta,\lambda,\phi_0,x_0,r_d)$. Here $H_0$ is the Hubble parameter,$\beta$ is the $f(Q,T)$ coupling parameter $\lambda$ characterizes the $\rm DBI$ warp factor, $\phi_0$ and $x_0$ represent the present values of the scalar fields and its velocity variable, respectively, and $r_d$ is the sound horizon at the drag epoch. we have chosen the scalar potential as $V(\phi)=\frac{1}{2}m^2\phi^2$, Here the parameter $m$ is not treated as an independent. We determined it from the cosmological constraint. At $z=0$ the dark-energy density satisfies 
\begin{equation}
    3H_0^2(1-\Omega_{m_0})=(\gamma_0-1)\lambda\phi^4_0+\frac{m^2\phi_0^2}{2}
\end{equation} 
using this relation we can obtain the value of $m$ as
\begin{equation}
    m=\sqrt{\frac{2[3H_0^2(1-\Omega_{m_0})-(\gamma_0-1)\lambda \phi^4_0]}{\phi_0^2}}
\end{equation}
this reduces the dimensionality of the parameter space and avoids introducing a redundant fitting parameter.

\begin{table*}
    \centering
    \begin{tabular}{|c|c|c|c|c|c|c|c|}\hline
         $\tilde{z}$&  \textbf{0.295}&  \textbf{0.510}&  \textbf{0.706}&  \textbf{0.934}&  \textbf{1.321}&  \textbf{1.484}& \textbf{2.330}\\\hline
         $D_V(\tilde{z})/r_d$&  7.942$\pm$0.075&  12.720$\pm$0.099&  16.050$\pm$0.110&  19.721$\pm$0.091&  24.252$\pm$0.174&  26.055$\pm$0.398& 31.267$\pm$0.256\\\hline
         $D_M(\tilde{z})/r_d$&  -&  13.588$\pm$0.167&  17.351$\pm$0.33&  21.576$\pm$0.152&  27.601$\pm$0.318&  30.512$\pm$0.760& 38.988$\pm$0.531\\\hline
         $D_H(\tilde{z})/r_d$&  -&  21.863$\pm$0.425&  19.455$\pm$0.330&  17.641$\pm$0.193&  14.176$\pm$0.221&  12.817$\pm$0.516& 8.632$\pm$0.101\\ \hline
    \end{tabular}
    \caption{DESIBAO measurements from the DESI Data Release~2 (DR2).}
    \label{tab:placeholder13}
\end{table*}

\section{Description of the figures and tables in the Appendix}
\label{appen:C}
Appendix~\ref{appen:C} summarizes the supplementary figures and tables that support the parameter estimation and model comparison presented in the main text. While the main text validates the model through a joint analysis of all three late-time data sets (cosmic chronometers $H(z)$, Pantheon+SHOES, and DESI BAO), here we provide additional results for different data combinations. Specifically, we present (i) the MCMC posterior distributions and parameter correlations inferred from various subsets of late--time probes, (ii) goodness-of-fit comparisons of the proposed $f(Q,T)+\mathrm{DBI}$ model with $\Lambda$CDM using expansion-rate and distance-modulus measurements, and (iii) a summary of the observational compilations and the corresponding best-fit constraints and statistical indicators. The items below describe each figure and table in this appendix.

\begin{figure}[H]
\centering
\includegraphics[width=7.9cm, height=7.8cm]{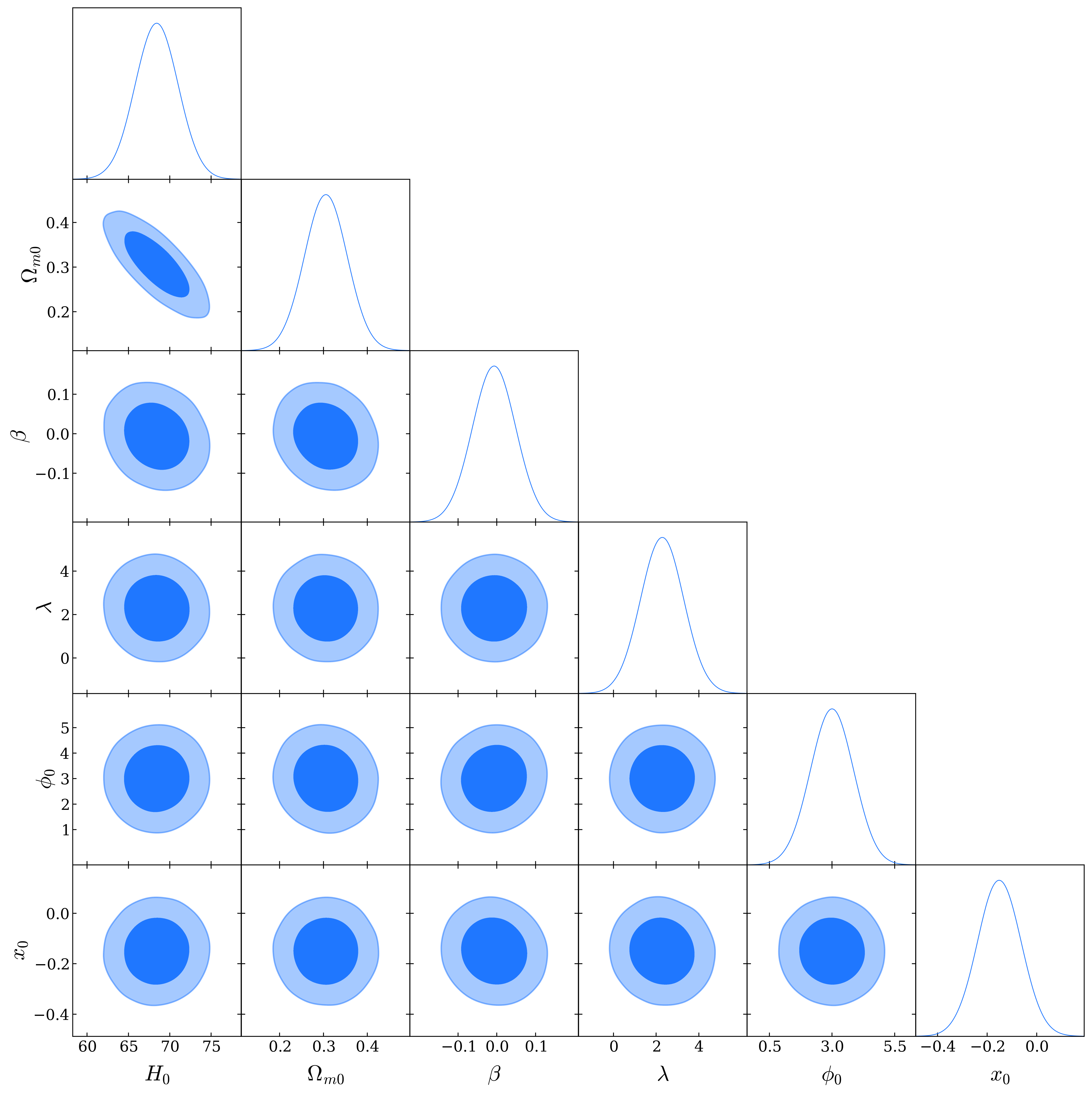}
\caption{Marginalized posterior distributions (1D) and joint credible regions (2D: 68\% and 95\%) for $H_0$, $\Omega_{m0}$, and $m$ inferred from the $H(z)$ dataset.}
\label{HT}
\end{figure}

\begin{figure}[H]
\centering
\includegraphics[width=7.9cm, height=7.0cm]{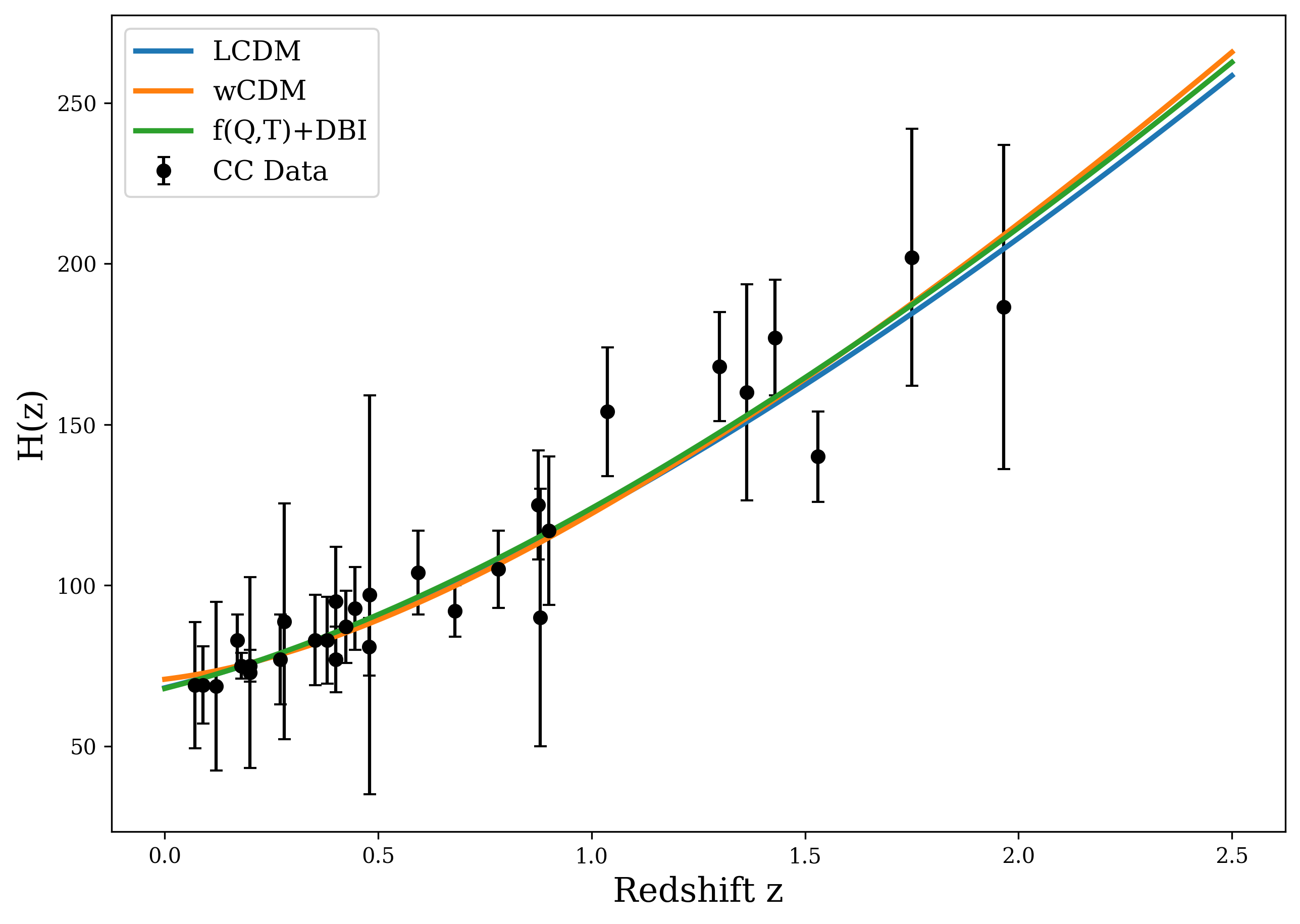}
\caption{Error-bar plot of the $H(z)$ measurements compared with the best-fit $f(Q,T)+\mathrm{DBI}$ model, as well as the best-fit $w$CDM and $\Lambda$CDM models.}
    \label{HE}
\end{figure}

\begin{figure}[H]
\centering
\includegraphics[width=7.9cm, height=7.8cm]{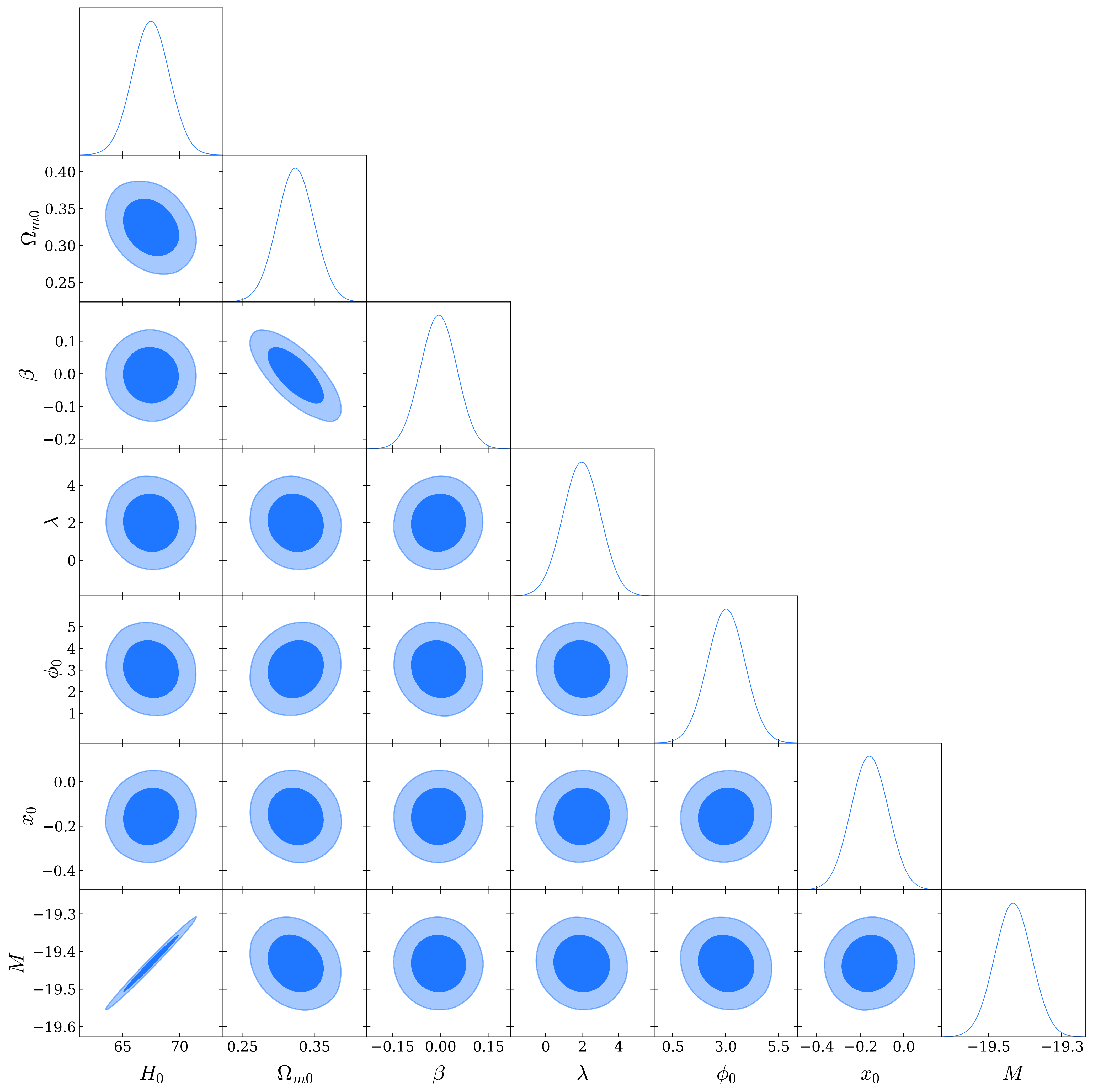}
\caption{Corner plot showing the 1D marginalized posteriors and 2D joint credible regions (68\% and 95\%) from the joint analysis of $H(z)$ and Pantheon+SHOES supernovae.}
    \label{PT}
\end{figure}

\begin{figure}[H]
\centering
\includegraphics[width=7.9cm, height=7.8cm]{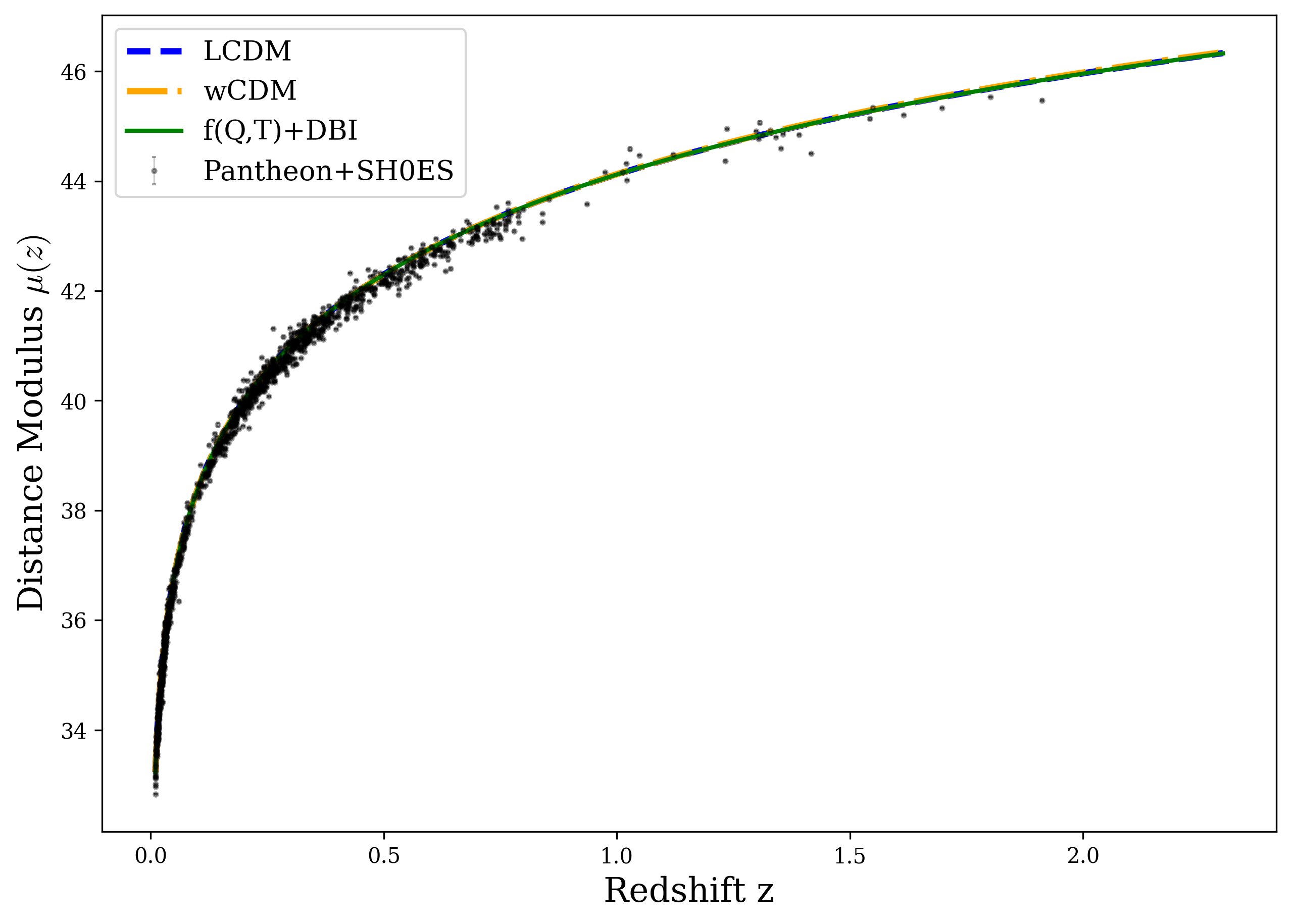}
\caption{Distance modulus $\mu(z)$ from the Pantheon+SHOES sample compared with the best-fit $f(Q,T)+\mathrm{DBI}$ model and the corresponding $\Lambda$CDM prediction. Error bars indicate the observational uncertainties.}
    \label{PE}
\end{figure}

\begin{figure}[H]
\centering
\includegraphics[width=7.9cm, height=7.8cm]{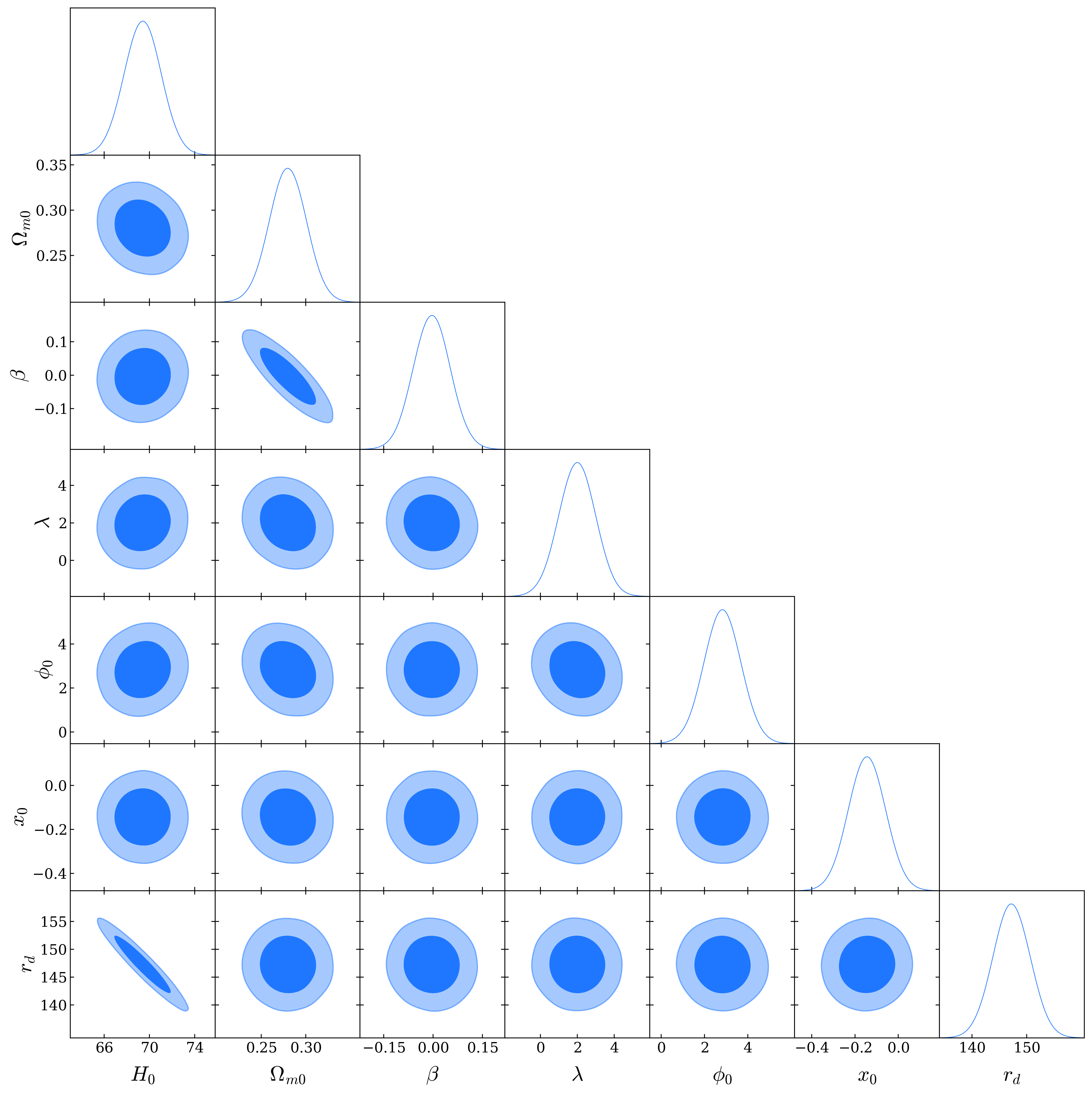}
\caption{Corner plot showing the 1D marginalized posteriors and 2D joint credible regions (68\% and 95\%) from the joint analysis of $H(z)$ and DESI BAO data.}
    \label{HDB}
\end{figure}

\begin{figure}[H]
\centering
\includegraphics[width=7.9cm, height=7.8cm]{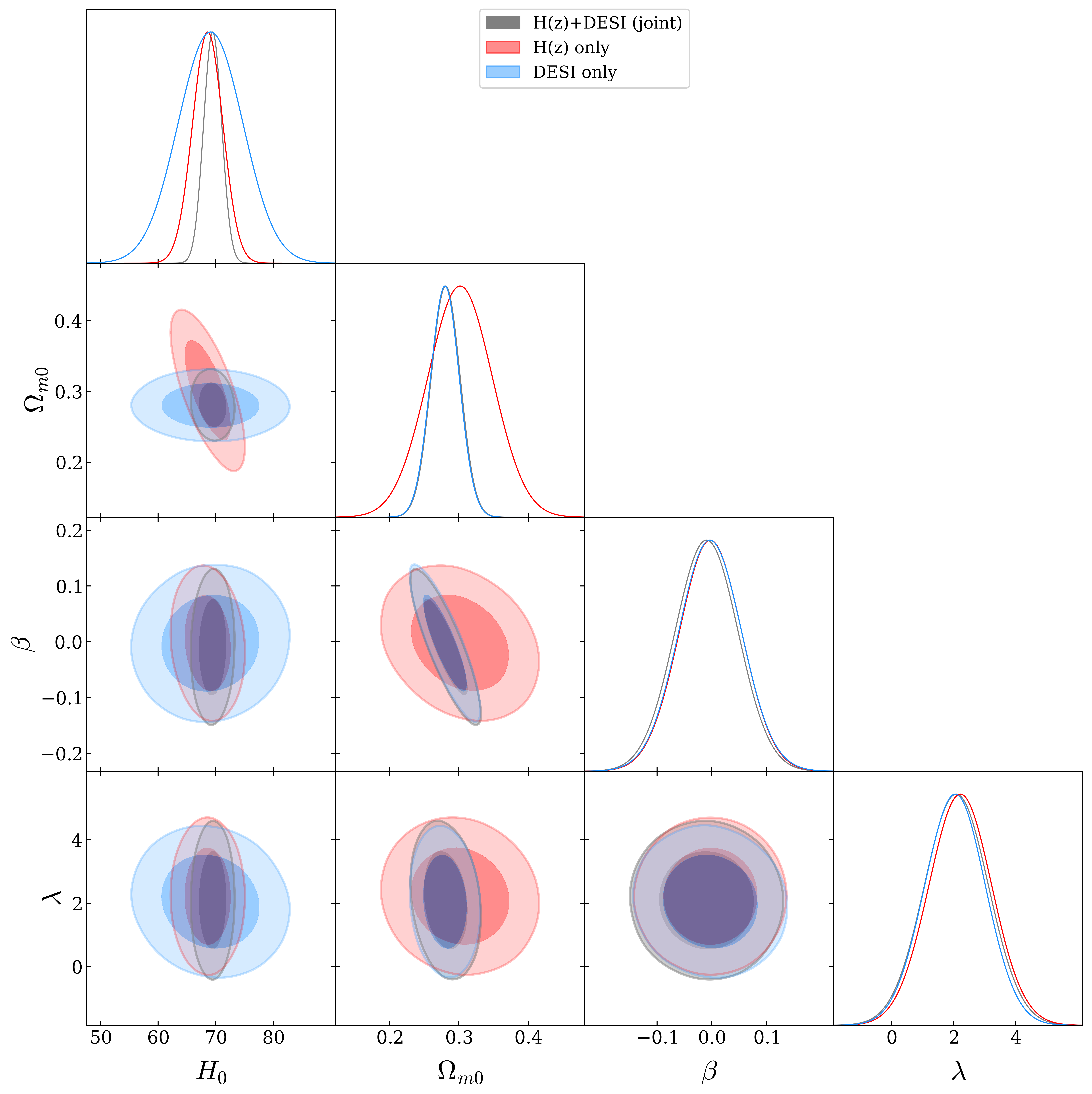}
\caption{Marginalized 1$\sigma$ and 2$\sigma$ confidence contours for the model parameters obtained from the $H(z)$-only, BAO-only, and joint $H(z)$+BAO analyses. The combined dataset reduces parameter degeneracies and tightens the constraints on $H_0$, $\Omega_{m0}$, $\beta$, and $\lambda$.}
    \label{fig8}
\end{figure}

\begin{figure}[H]
\centering
\includegraphics[width=7.9cm, height=7.8cm]{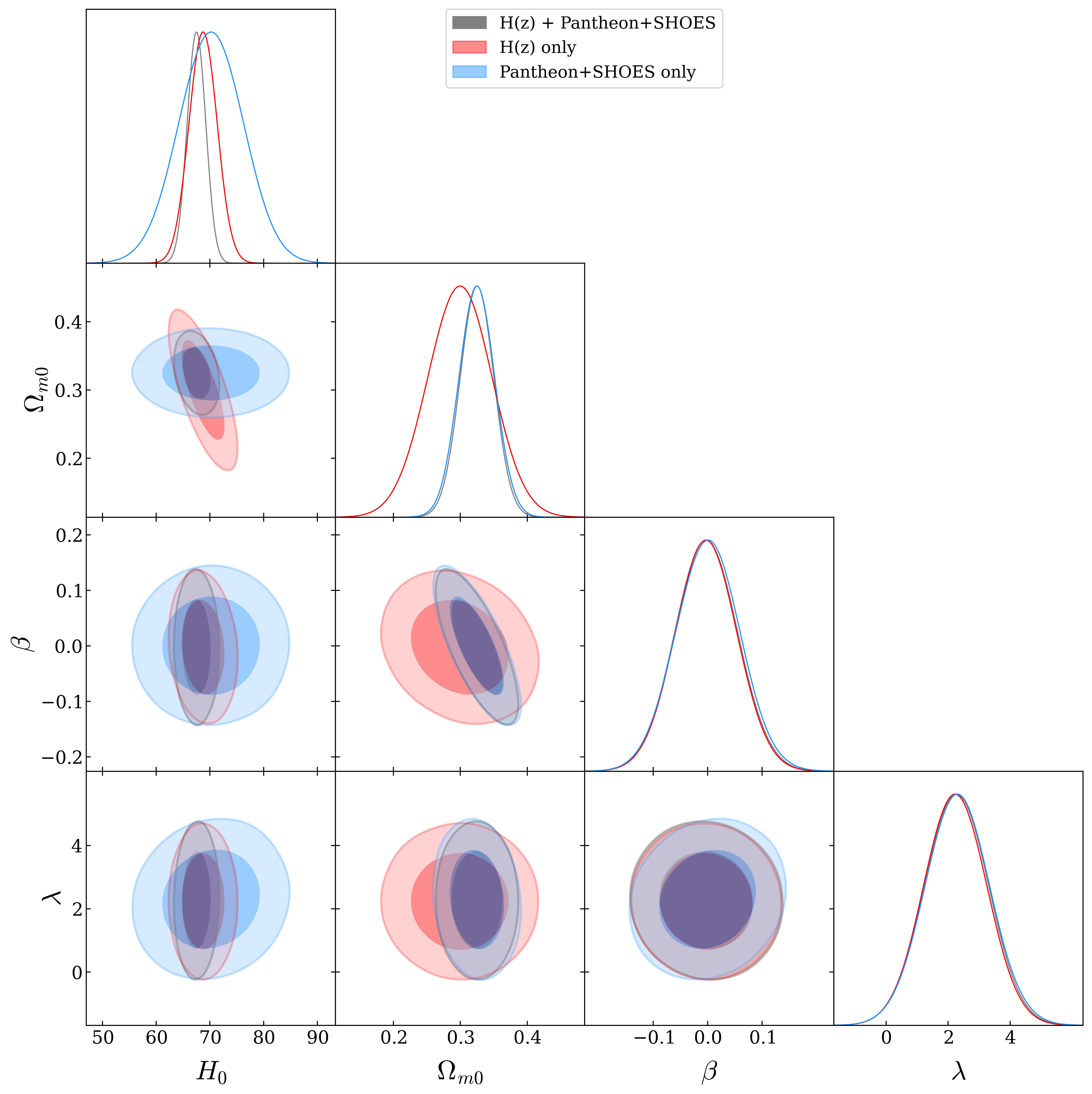}
\caption{Marginalized 1$\sigma$ and 2$\sigma$ confidence contours for the model parameters obtained from the $H(z)$-only, Pantheon+SHOES-only, and joint $H(z)$+Pantheon+SHOES analyses. The combined dataset reduces parameter degeneracies and tightens the constraints on $H_0$, $\Omega_{m0}$, $\beta$, and $\lambda$.}
    \label{fig9}
\end{figure}

\begin{table*}[t]
    \centering
    \renewcommand{\arraystretch}{1.3}
\begin{tabular}{c c c c c c c c}
\hline\hline
Data & Parameter & Best-fit Value & $\chi^2_{\text{red}}$ & AIC & BIC & $\Delta$AIC & $\Delta$BIC \\
\hline
\multirow{2}{*}{H(z)} 
  & $H_0$     &  68.864$\pm$2.618 &  &  &  &  &  \\ 
  & $\Omega_{m_0}$     &  0.280$\pm$0.049 &  &  &  &  &  \\ & $\beta$     & 0.025$\pm$0.056  & 0.626 & 28.29 & 37.08 & 6.96 & 12.82 \\ & $\lambda$     & 3.861$\pm$1.011 &  &  &  &  &  \\ & $\phi_0$     & 4.433$\pm$0.864 &  &  &  &  &  \\ & $x_0$     & -0.103$\pm$0.088 &  &  &  &  &  \\ 
\hline
\multirow{4}{*}{H(z)+Patheon+SHOES} 
  & $H_0$     & 67.781$\pm$1.610  &  &  &  &  &  \\ 
  & $\Omega_{m_0}$     & 0.324$\pm$0.026  & &  &  &   &  \\ & $\beta$     & 0.005$\pm$0.057  & 0.881 & 1435.55 & 1473.28 & 7.18 & 28.74 \\ 
& $\lambda$     & 0.637$\pm$1.021  &  &  &  &  &  \\ & $\phi_0$     & 1.661$\pm$0.883 &  &  &  &  &  \\ & $x_0$     & -0.250$\pm$0.085 &  &  &  &  &  \\ & $M$     & -19.422$\pm$0.050 &  &  &  &  &  \\ 
\hline
\multirow{4}{*}{H(z)+ DESIBAO} 
  & $H_0$     & 69.137$\pm$1.641  &  &  &  &  &  \\ 
  & $\Omega_{m_0}$     & 0.294$\pm$0.021  &  &  &  &   &  \\ & $\beta$     & -0.005$\pm$0.056  & 0.925 & 54.68 & 68.21 & 7.52 & 15.25 \\ 
 & $\lambda$& 
 0.546$\pm$0.999 &  &  &  &  &  \\
  & $\phi_0$& 
 1.831$\pm$0.858 &  &  &  &  &  \\  & $x_0$& 
 -0.011$\pm$0.086 &  &  &  &  &  \\  & $r_d$& 
 146.921$\pm$3.407 &  &  &  &  &  \\ 
\hline
\end{tabular}
\caption{Summary of Best-Fit Parameters and Model Comparison Statistics}
\label{stat}
\end{table*}

\begin{itemize}
\item Fig.~\ref{HT} presents the triangle plot of the posterior distributions of the model parameters $H_0$, $\Omega_{m0}$, $\beta$, $\lambda,\phi_0$ and $x_0$ inferred from the $H(z)$ (cosmic-chronometer) data. The one-dimensional marginalized posteriors are smooth and close to Gaussian, indicating stable parameter estimation and good MCMC convergence. Overall, the $H(z)$ dataset provides meaningful constraints on the proposed model and is consistent with the observed late-time expansion history. Fig.~\ref{HE} compares the observed Hubble-parameter measurements with the best-fit predictions of the proposed $f(Q,T)+\mathrm{DBI}$ model and the standard $\Lambda$CDM scenario. The $f(Q,T)+\mathrm{DBI}$ model closely tracks the $H(z)$ data across the full redshift range and remains within the observational uncertainties, demonstrating an excellent fit to late-time expansion-rate measurements.

\item Fig.~\ref{PT} shows the triangle plot of the posterior distributions of $H_0$, $\Omega_{m0}$, $\beta$, $\lambda, \phi_0, x_0$ and $M$ obtained from the joint analysis of $H(z)$ and Pantheon+SHOES Type~Ia supernova data. The marginalized posteriors are smooth and unimodal, supporting good convergence of the MCMC sampling. The $68\%$ and $95\%$ confidence regions reveal the expected correlations among parameters, and the combined dataset yields tighter and more robust constraints than those obtained from either probe alone. Fig.~\ref{PE} displays the supernova Hubble diagram, i.e., the observed distance modulus $\mu(z)$ as a function of redshift, compared with the best-fit predictions of the $f(Q,T)+\mathrm{DBI}$ model and $\Lambda$CDM. Both models provide an excellent description of the supernova data across the covered redshift range, with any differences between their late-time expansion histories becoming most apparent at higher redshift.

\item Fig.~\ref{HDB} presents the marginalized posterior distributions and confidence contours obtained from the joint $H(z)$+DESI BAO analysis. The inclusion of DESI BAO measurements significantly improves the constraints on the cosmological parameters and on the sound-horizon scale $r_d$, illustrating the complementarity between expansion-rate data and standard-ruler information. Fig.~\ref{fig8} compares the $1\sigma$ and $2\sigma$ confidence contours from $H(z)$ only, DESI BAO only, and the combined $H(z)$+DESI BAO analysis. The joint dataset reduces parameter degeneracies and yields tighter constraints on $H_0$, $\Omega_{m0}$, $\beta$ and $\lambda$. Fig.~\ref{fig9} shows the corresponding confidence contours for the $H(z)$ only, Pantheon+SHOES only, and joint $H(z)$+Pantheon+SHOES analyses, demonstrating that the addition of supernova luminosity-distance information further tightens the allowed parameter space.

\item Table~\ref{stat} summarizes the results shown in Figs.~\ref{HT}--\ref{fig9} for all parameter combinations. Overall, the fitted model and nuisance parameters are well constrained and broadly consistent with current observational values.
\end{itemize}

\section{Additional Remarks on the Observational Analysis}
\label{appen:D}
The present investigation is motivated by the ongoing search for viable alternatives to the standard cosmological paradigm that can accommodate the observed late-time acceleration of the Universe while providing insights into the nature of dark energy and modified gravity. Our observational analysis demonstrates that the proposed $f(Q,T)$ gravity model in the presence of a DBI-essence scalar field offers a consistent description of the late-time cosmic expansion when constrained by the combined Hubble, Pantheon+SHOES, and DESI BAO datasets. The inferred matter density parameter, $\Omega_{m0}=0.319\pm0.020$, is in excellent agreement with values obtained from contemporary low-redshift observations, while the matter--geometry coupling parameter, $\beta=-0.070\pm0.057$, remains statistically compatible with zero, indicating that current data favor a regime close to the symmetric teleparallel equivalent of General Relativity. The DBI-sector parameters are constrained with moderate precision, reflecting the sensitivity of present observations to scalar-field dynamics. Furthermore, the best-fit value of the Hubble constant, $H_0=69.231\pm1.738,\mathrm{km,s^{-1},Mpc^{-1}}$, is closer to local distance-ladder measurements than to CMB-based $\Lambda$CDM estimates, suggesting that the additional degrees of freedom associated with the DBI scalar field and the modified-gravity sector can modify the late-time expansion history and potentially alleviate the Hubble-tension problem while maintaining a statistically competitive fit to the data. The statistical methodology adopted here is sufficiently general to be applied to a wide range of modified-gravity and dynamical dark-energy models using both current and forthcoming cosmological observations. Consequently, the framework developed in this work provides a valuable tool for assessing the observational viability of alternative gravitational theories and for probing the fundamental physics driving cosmic acceleration in the era of precision cosmology.

\end{document}